\documentclass[11pt,a4paper]{article}
\usepackage{jheppub}
\usepackage[utf8]{inputenc}
\usepackage[english]{babel}
\usepackage{amsfonts}
\usepackage{tcolorbox}[most] 

\usepackage[bottom]{footmisc}

\usepackage [autostyle]{csquotes}
\MakeOuterQuote{"}

\newcommand{\sgn}[1]{\mathrm{sgn}\!\left(#1\right)}
\newcommand{\sgnsi}{\mathrm{sgn}_{123}}

\newcommand{\Li}[2]{\mathrm{Li}_{#1}\!\left(#2\right)}
\newcommand{\Fpm}[1]{\mathcal{F}_{\pm}\!\left(\eps; #1\right)\!}
\newcommand{\Fmp}[1]{\mathcal{F}_{\mp}\!\left(\eps; #1\right)\!}
\newcommand{\Fp}[1]{\mathcal{F}_{+}\!\left(\eps; #1\right)\!}
\newcommand{\LinSV}[1]{\mathcal{L}_{#1}}
\newcommand{\LiSV}[2]{\mathcal{L}_{#1}\!\left(#2\right)}
\newcommand{\LiSVLewin}[2]{\mathrm{L}_{#1}\!\left(#2\right)}
\newcommand{\LiSVRamakrishnan}[2]{\mathrm{D}_{#1}\!\left(#2\right)}
\newcommand{\LiSVZagier}[2]{\mathrm{P}_{#1}\!\left(#2\right)}

\newcommand{\eps}{\varepsilon}
\newcommand{\dx}{\mathrm{d}}
\newcommand{\e}{\mathrm{e}}
\newcommand{\iu}{\mathrm{i}}
\newcommand{\iz}{\mathrm{i}0}
\newcommand{\iztilde}{\mathrm{i}\tilde{0}}

\newcommand{\BernoulliB}[1]{\mathrm{B}_{#1}}
\newcommand{\hyg}[1]{{_2}\mathrm{F}_1\!\left(#1\right)}
\newcommand{\Fbox}[2]{\mathcal{F}_{#1}\!\left(#2\right)}
    \DeclareFontFamily{U}{wncy}{}
    \DeclareFontShape{U}{wncy}{m}{n}{<->wncyr10}{}
    \DeclareSymbolFont{mcy}{U}{wncy}{m}{n}
    \DeclareMathSymbol{\Sh}{\mathord}{mcy}{"58}

\title{The  massless single off-shell scalar box integral --- branch cut structure and all-order epsilon expansion}

\author[]{Juliane Haug}
\author[1]{and Fabian Wunder\note{Corresponding author}}

\affiliation[]{Institut f\"ur Theoretische Physik, Universit\"at T\"ubingen, \\
	Kepler Center for Astro and Particle Physics, \\ 
	Auf der Morgenstelle 14, D-72076 T\"ubingen, Germany}

\emailAdd{juliane-clara-celine.haug@uni-tuebingen.de}
\emailAdd{fabian.wunder@uni-tuebingen.de}

\abstract{
We investigate the single off-shell scalar box integral with massless internal lines in dimensional regularization. A special emphasis is given to higher orders in the dimensional regularization parameter epsilon, its branch cut structure, and kinematic limits.
Common representations of the box integral introduce superficial branch cuts, which we eliminate to all orders in the epsilon expansion. In the literature so far a satisfactory solution for this issue only exists up to finite order in the epsilon expansion. However, for certain calculations at NNLO in perturbation theory, higher orders in epsilon of this integral are required. 
In this paper, we present results to all orders in epsilon in terms of \textit{single-valued polylogarithms} and explicitly determine the real and imaginary part of the box integral in all kinematic regions.}
\keywords{Feynman integrals, perturbative QCD, dimensional regularization, 
epsilon expansion}
\arxivnumber{2211.14110}

\begin{document}
\maketitle

\flushbottom
                                   
\clearpage 

\section{Introduction}	
With the upcoming electron-ion collider (EIC) \citep{Accardi:2012qut}, it is essential to evaluate the hard-scattering coefficients for semi-inclusive deep-inelastic scattering (SIDIS) at next-to-next-to-leading order (NNLO) in QCD \citep{Anderle:2016kwa,Abele:2021nyo,Borsa:2022vvp,Abele:2022wuy}. One type of contributions arising at NNLO are the one-loop corrections to the next-to-leading-order processes of real gluon emission and photon-gluon fusion. One-loop integrals such as these arise in virtually all higher-order perturbative calculations and were thoroughly studied in the past 40 years \citep{tHooft1979,vanNeerven1984,vanOldenborgh1990,Beenakker1990,Denner1991,vanOldenborgh1991,
Hahn1999,Fleischer2003,Patel2015,Carrazza2016,FabiValery}. Given that tensorial loop integrals can be systematically reduced to scalar integrals using the Passarino-Veltman technique \citep{PassarinoVeltman}, it is sufficient to evaluate scalar integrals. The case of massless propagators is of particular interest for perturbative QCD calculations, since light quark masses can be neglected in high-energy scattering processes and gluons are massless. Loop integrals with vanishing internal masses generally have soft and collinear divergences, which must be regularized. The most commonly used method is dimensional regularization, which simultaneously serves to regularize ultraviolet and infrared divergences \citep{Bollini:1972ui,tHooft:1973mfk,Gastmans:1973uv,Leibbrandt:1975dj}. We will work in $d=4-2\eps$ dimensions.

Both real and imaginary part of a loop integral do in general yield contributions to the absolute value squared of a matrix element, hence we must know both. The imaginary part of a loop integral arises in certain kinematic regions and will be governed by the causal $+\iz$ from the propagators. For massless two- and three-propagator scalar loop integrals in general dimensions $d$, it is easy to keep track of this infinitesimal imaginary part.\footnote{By dimensional analysis, massless one-propagator tadpole integrals must vanish since they depend on no mass scale.} Following the calculation given in ref.\,\citep{FabiValery}, one finds that the external momentum invariants are essentially replaced by $(p_i\pm p_j)^2\rightarrow (p_i\pm p_j)^2 +\iz$ in the final results. This simple rule yields no unambiguously well-defined imaginary part for the four-propagator box integral in general dimension $d$, as we will discuss in section \ref{sec:Calculating_D0} below.

The scalar box integral with one off-shell external particle (i.e. with non-vanishing momentum squared) is especially important since it is needed for higher order (NNLO) perturbative calculations of processes such as SIDIS, Drell-Yan, or single inclusive annihilation (SIA), which all feature a single off-shell gauge boson. In ref.\,\citep{FabriciusSchmitt}, this integral and its imaginary part were first given up to finite order $\eps^0$ in dimensional regularization. The result for general dimensions $d$ written in terms of three Gauss hypergeometric functions was obtained in refs.\,\citep{Matsuura1989,BernDixonKosower1}. However, the simple prescription $(p_i\pm p_j)^2\rightarrow (p_i\pm p_j)^2 +\iz$ given in ref{.}\,\citep{BernDixonKosower1} for analytic continuation of the result to all kinematic regions only applies to the expansion up to finite order. Ref{.}\,\citep[eq{.}\,(D.4)]{Matsuura1989}, which was used in the calculation of DIS structure functions at NNLO \citep{Zijlstra1992}, does not touch on the question of analytic continuation. In \citep{FabiValery}, the box integral was expanded to all orders in $\eps$, however the branch cut structure was also not discussed. The causal $+\iz$ from the propagators was systematically kept in ref.\,\citep{DuplancicNizic}, but the result was only given up to finite order in dimensional regularization. To regularize divergences occurring in kinematic limits, however, we need the full $\eps$-dependence of any terms divergent in these kinematic limits.

In this paper, we independently recalculate the massless single off-shell scalar box integral for general dimensions $d$. We explicitly keep the causal $+\iz$ from the propagators, such that our results are valid for arbitrary (real) values of the kinematic variables. This allows us to explicitly determine real and imaginary part to all orders of the $\eps$-expansion, which goes beyond the finite $\mathcal{O}(\eps^0)$ results of refs.\,\citep{BernDixonKosower1,DuplancicNizic} and the \textit{Mathematica} implementation in \textit{Package}-\texttt{X} \citep{Patel2015}. In section \ref{sec:Calculating_D0}, we calculate the internal-massless single-off shell scalar box integral using the well-known Feynman parametrization approach and an elegant factorization which allows for a trivial reduction to one-dimensional integrals. As a limiting case our result also includes the internal-massless box integral with on-shell massless external particles, which is needed for direct photon production and Compton-like scattering processes. In section \ref{sec:Epsilon expansion}, we perform the $\eps$-expansion to all orders and the explicit determination of real and imaginary parts. The special cases of vanishing Mandelstam variables, as well as several identities for the Gauss hypergeometric function, and the definition of \textit{single-valued polylogarithms}, are collected in the appendices. Additionally, we provide an example illustrating the need of higher orders in $\eps$ of the scalar box integral for phase space regularization in appendix \ref{app:Regularizing_kinematic_limits}.

\section{Calculating the massless single off-shell scalar box integral}\label{sec:Calculating_D0}
	\begin{figure}[b]
		\centering
			\includegraphics[width=0.4\textwidth]{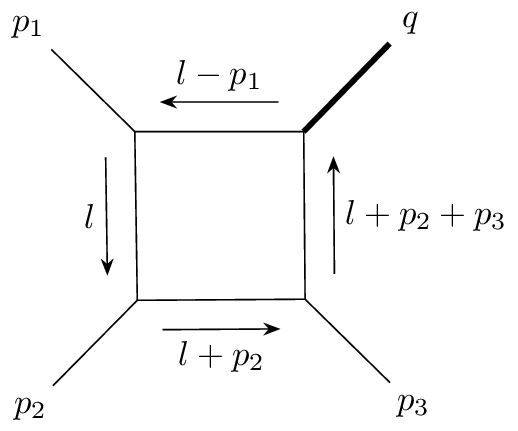}
		\caption{The massless scalar box diagram with one off-shell external particle carrying momentum $q$. It is $p_1^2=p_2^2=p_3^2=0$ and $q^2\neq 0$. All external momenta are taken to be incoming. Drawn with Ti\textit{k}Z-Feynman \citep{Ellis2017}.}\label{fig:scalar_box_diagram}
	\end{figure}
\noindent{}The massless single off-shell scalar box integral is defined through the loop integral
	\begin{align}
		\mathrm{D}_0 \,\equiv\, \frac{\mu^{4-d}}{\iu \pi^{d/2}} \int\dx^dl\,
		\frac{1}{\left[l^2 + \iu 
		 0\right] \left[(l+p_2)^2 +\iz\right] \left[(l+p_2+p_3)^2 +\iz\right] \left[(l-p_1)^2 +\iz\right]} \,,
	\end{align}
where the external momenta are labelled as indicated by the Feynman diagram depicted in figure \ref{fig:scalar_box_diagram}. It is $p_1^2=p_2^2=p_3^2=0$ and $q^2=(p_1+p_2+p_3)^2\neq 0$. Note that we take all external momenta to be incoming. We have also explicitly kept the causal $+\iz$ in the propagators, which is necessary to determine the physical side of the branch cuts that appear in the final result.\footnote{Here, $\iz$, with $0$ taken to be positive, is an infinitesimal imaginary part indicating on which side of the branch cut a multivalued function should be evaluated.}
The prefactor is chosen as in ref.\,\citep{Ellis}.

To evaluate this loop integral, we first combine the propagators into one generalized propagator of higher power using the well-known Feynman parametrization \citep[eq.\,(A.39)]{Peskin1995}
	\begin{equation}
		\frac{1}{a b c d} \,= \, 6 \int_0^1\dx x_1\, \dx x_2\, \dx x_3\, \dx x_4\, \frac{\delta (1-x_1-x_2-x_3-x_4)}{\left[a x_1 \,+\, b x_2 \,+\, c x_3 \,+\, d x_4\right]^4} \,.
	\end{equation}
Next, we complete the square with respect to $l$ and then shift the loop momentum in the usual manner, which leads to
	\begin{align}
		\mathrm{D}_0 \,&=\, \mu^{4-d} \int_0^1\dx x_1\, \dx x_2\, \dx x_3\, \dx x_4\, \delta(1-x_1-x_2-x_3-x_4)
		\nonumber \\
		&\hphantom{=}\, \times\, \frac{1}{\iu \pi^{d/2}} \int\dx^dl' \frac{6}{\left[l'^2 + s_1 x_1(x_2+x_3) + s_2x_3(x_1+x_4) +s_3x_1x_3 + \iz\right]^4} \,, \label{eq:D0_l'_integral}
	\end{align}
where we introduced the Mandelstam variables
	\begin{align}
		s_1 \,&=\, (p_1+p_2)^2 \,, \\
		s_2 \,&=\, (p_2+p_3)^2 \,, \\
		s_3 \,&=\, (p_1+p_3)^2 \,.
	\end{align}
In the following, we treat all Mandelstam variables as real, but otherwise arbitrary.

Evaluating the loop integral on the second line of eq.\,\eqref{eq:D0_l'_integral} with the help of \citep[eq.\,(A.44)]{Peskin1995}
	\begin{equation}
		\frac{1}{\iu \pi^{d/2}} \int\dx^dl\, \frac{1}{\left[l^2-\Delta + \iz\right]^{\alpha}} \,=\, (-1)^{\alpha} \frac{\Gamma(\alpha-d/2)}{\Gamma(\alpha)}(\Delta - \iz)^{d/2-\alpha}
	\end{equation}
results in
	\begin{align}
		D_0 \,&=\,\mu^{2\eps}\,\Gamma(2+\eps)\int_0^1 \frac{\dx x_1\, \dx x_2\, \dx x_3\, \dx x_4\,\delta(1-x_1-x_2-x_3-x_4)}{\left[-s_1 x_1(x_2+x_3) - s_2x_3(x_1+x_4) - s_3x_1x_3 - \iz\right]^{2+\eps}} \,.
		\label{eq:D0_FeynmanParametrized}
	\end{align}
Note that this expression is symmetric under exchanging $s_1 \leftrightarrow s_2$. To reflect this, we write the scalar box integral as $D_0(s_1,s_2,q^2)$, where $q^2=s_1+s_2+s_3$ is the squared momentum of the off-shell external particle. To decouple the Feynman parameter integrals, which are currently coupled via the delta function, we use the substitution
	\begin{align}
		x_1 \,&=\, \eta_1\xi_1\,, \quad x_4 \,=\, \eta_1(1-\xi_1)\,, \\
		x_3 \,&=\, \eta_2\xi_2\,, \quad x_2\,=\, \eta_2(1-\xi_2)\,.
	\end{align}
The Jacobian of this substitution, which was inspired by ref.\,\citep[page 44]{Smirnov_Feynman_Integrals}, is $\eta_1\eta_2$. Through the substitution, the $\eta$- and $\xi$-integrals factorize,
	\begin{align}
		D_0\!\left(s_1,s_2,q^2\right) \,&=\, \mu^{2\eps}\, \Gamma(2+\eps) \int_0^1\dx\eta_1 \int_0^1\dx\eta_2\; \eta_1^{-\eps-1}\, \eta_2^{-\eps-1}\, \delta(1-\eta_1 - \eta_2) \nonumber
		\\
		&\hphantom{=}\,\times \int_0^1\dx\xi_1 \int_0^1\dx\xi_2\, \left[- s_1 \xi_1 - s_2\xi_2 - s_3\xi_1\xi_2 - \iz \right]^{-\eps-2}
		\nonumber
		\\
		&=\, \mu^{2\eps}\, \frac{\Gamma(2+\eps) \Gamma^2(-\eps)}{\Gamma(-2\eps)} \int_0^1\dx\xi_1 \int_0^1\dx\xi_2\, \left[- s_1 \xi_1 - s_2\xi_2 - s_3\xi_1\xi_2 -\iz\right]^{-\eps-2} \,. \label{eq:D0_xi_integrals}
	\end{align}
Here, the delta function was used to evaluate one of the $\eta$-integrals. The remaining \mbox{$\eta$-integral} thus became a representation of the Beta function, which was evaluated in terms of Gamma functions,
	\begin{equation}
		B(a,b) \,\equiv\, \int_0^1\dx t\, t^{a-1} (1-t)^{b-1} \,=\, \frac{\Gamma(a)\, \Gamma(b)}{\Gamma(a+b)} \,.
	\end{equation}
For the resulting integral to safely converge, we must take $\eps < 0$.\footnote{In the special cases of one or two (or three) vanishing Mandelstam variables $\eps<-1$ $(\eps < -2)$ is required.}
We will first evaluate the integral in this area of convergence and subsequently analytically continue the end result to larger $\eps$.
% This is the only way to define the expansion around $\eps=0$ for this integral that does not converge for $\eps=0$! Think of it as a calculation prescription in dimensional regularization.

So far, our calculation did not differ much from the way the scalar box integral was calculated before (see e{.}\,g{.} ref.\,\citep{DuplancicNizic}) and was mainly presented for pedagogical reasons and introduction of notation. We will now deviate from the usual path by factoring the term $\frac{s_1 s_2}{s_3}$ out of the integrand. This is of course not sensible for ${\frac{s_1 s_2}{s_3} = 0, \pm\infty}$, that is $s_1, s_2, s_3 = 0$. The special cases of one or several vanishing Mandelstam variables will be considered in appendix \ref{app:Vansishing_Mandelstam_variables}. Using the identity (compare \citep[eq.\,(19)]{DuplancicNizic})
	\begin{align} 
		(a-\iz)^\alpha \,=\, (b-\iz)^\alpha \left(\frac{a}{b}-\iz\, \sgn{b}\right)^\alpha ,\quad \text{where } a\in \mathbb{R}\,,\; b\in\mathbb{R}\!\setminus\! \lbrace 0\rbrace \,,\; \alpha\in\mathbb{C}\,,
		\label{eq:identity_for_factoring}
	\end{align}
which is not completely trivial since $\alpha$ is not an integer, we obtain
	\begin{align}
		D_0\!\left(s_1,s_2,q^2\right)&=\,\mu^{2\eps}\, \frac{\Gamma(2+\eps) \Gamma^2(-\eps)}{\Gamma(-2\eps)} \left(\frac{s_1s_2}{s_3} - \iz\right)^{\!-\eps-2} \nonumber \\
		&\hphantom{=}\times \int_0^1\dx\xi_1 \int_0^1\dx\xi_2\, \left[-\frac{s_3}{s_2} \xi_1 -\frac{s_3}{s_1} \xi_2 -\frac{s_3^2}{s_1s_2}\xi_1\xi_2 -\iz\, \sgn{\frac{s_3}{s_1s_2}}\right]^{-\eps-2} \,.
	\end{align}
The $\iz$ prescription determines on which side of the branch cut the respective factors should be evaluated if their argument is negative. The remaining integral on the second line only depends on the two dimensionless variables
	\begin{equation}
		x_1 \,\equiv\, -\frac{s_3}{s_1}\,, \quad x_2 \,\equiv\, -\frac{s_3}{s_2} \,,
		\label{eq:Def_x1x2}
	\end{equation}
which is why factoring the term $\frac{s_1 s_2}{s_3}$ out of the integrand is advantageous. By substituting $\xi_1\rightarrow\zeta_2=x_2\xi_1$ and $\xi_2\rightarrow\zeta_1= x_1\xi_2$,
all dependence on the dimensionless $x_i$ is moved into the integration boundaries,
	\begin{align}
		D_0\!\left(s_1,s_2,q^2\right) &=\, \frac{\Gamma(2+\eps) \Gamma^2(-\eps)}{\Gamma(-2\eps)}\, \frac{1}{s_1s_2} \left(\frac{s_3\mu^2}{s_1s_2} + \iz\right)^{\!\eps} I_{12}(x_1,x_2)\,,\\
		\text{where}\quad I_{12}(x_1,x_2)\,&\equiv\,\int_0^{x_1}\dx\zeta_1 \int_0^{x_2}\dx\zeta_2\, \left[1-(1-\zeta_1)(1-\zeta_2) -\iz\,\sgnsi \right]^{-\eps-2} \,.
	\end{align}
Note that we inverted the prefactor using
	\begin{equation}
		(a \pm \iz)^\alpha \,=\, \left(\frac{1}{a} \mp \iz\right)^{\!-\alpha} ,\quad \text{where } a\in \mathbb{R}\!\setminus\! \lbrace 0\rbrace \,. \label{eq:identity_for_inverting}
	\end{equation}
Additionally, we introduced the abbreviation
	\begin{align}
		\sgnsi \,\equiv\, \sgn{\frac{s_3}{s_1s_2}}.
		\label{eq:Def_sgnsi}
	\end{align}
And furthermore, we have written the integrand in a form suggesting the substitution $\zeta_i \rightarrow 1-\zeta_i$. Subsequently substituting $\zeta_2 \rightarrow \zeta_{12} = 1-\zeta_1\zeta_2$ allows us to evaluate the resulting $\zeta_{12}$ integral,
	\begin{align}
		&I_{12}(x_1,x_2)\,=\,\int_1^{1-x_1}\dx\zeta_1 \int_1^{1-x_2}\dx\zeta_2 \left[1- \zeta_1\zeta_2 -\iz\, \sgnsi\right]^{-\eps-2} \nonumber
		\\
		&=\, - \int_1^{1-x_1}\frac{\dx\zeta_1}{\zeta_1} \int_{1-\zeta_1}^{1-\zeta_1(1-x_2)}\dx\zeta_{12} \left[\zeta_{12} -\iz\, \sgnsi\right]^{-\eps-2} \nonumber
		\\
		&=\, \frac{1}{1+\eps}\int_1^{1-x_1}\frac{\dx\zeta_1}{\zeta_1} \left\lbrace \left[1-\zeta_1(1-x_2)-\iz\, \sgnsi\right]^{-\eps-1}- \left[1-\zeta_1-\iz\, \sgnsi \right]^{-\eps-1} \right\rbrace .
	\end{align}
While this integrand is finite at $\zeta_1 = 0$, we must regularize the divergence at $\zeta_1 = 0$ if we split the integral into two separate integrals. For this, we subtract and add $1$ and subsequently substitute $\zeta_1 \rightarrow \zeta = 1-\zeta_1(1-x_2)$ in the first integral and $\zeta_1 \rightarrow \zeta = 1- \zeta_1$ in the second integral,
	\begin{align}
		I_{12}(x_1,x_2)\,&=\,\frac{1}{1+\eps} \left\lbrace \int_1^{1-x_1}\frac{\dx\zeta_1}{\zeta_1} \left(\left[1-\zeta_1(1-x_2)-\iz\, \sgnsi\right]^{-\eps-1} -\, 1\right) \right. \nonumber
		\\
		&\phantom{=\,\frac{1}{1+\eps}\lbrace}\left. - \int_1^{1-x_1}\frac{\dx\zeta_1}{\zeta_1}\left(\left[1-\zeta_1-\iz\, \sgnsi\right]^{-\eps-1} -\, 1\right) \right\rbrace \nonumber
		\\
		&=\, \frac{1}{1+\eps} \left\lbrace -\int_{x_2}^{1-(1-x_1)(1-x_2)}\frac{\dx\zeta}{1-\zeta} \left(\left[\zeta-\iz\, \sgnsi\right]^{-\eps-1} -\, 1 \right) \right. \nonumber
		\\
		&\phantom{=\, \frac{1}{1+\eps}\lbrace}\left. + \int_0^{x_1}\frac{\dx\zeta}{1-\zeta} \left(\left[\zeta - \iz\, \sgnsi\right]^{-\eps-1} -1 \right) \right\rbrace .
	\end{align}
Splitting the first of these integrals in two, we find that the scalar box integral is given by
	\begin{align}
		D_0\!\left(s_1,s_2,q^2\right) =\, &-\frac{1}{\eps} \frac{\Gamma(1+\eps)\, \Gamma^2(1-\eps)}{\Gamma(1-2\eps)}\, \frac{2}{s_1s_2} \left(\frac{s_3\mu^2}{s_1s_2} + \iz\right)^{\eps} \nonumber
		\\
		&\times \left\lbrace \int_0^{x_1}\frac{\dx\zeta}{1-\zeta} \left(\left[\zeta - \iz\, \sgnsi\right]^{-\eps-1} -1 \right) \right. \nonumber
		\\
		&\hphantom{\times\lbrace}\;\;+\int_0^{x_2}\frac{\dx\zeta}{1-\zeta} \left(\left[\zeta - \iz\, \sgnsi\right]^{-\eps-1} -1 \right) \nonumber
		\\
		&\hphantom{\times\lbrace}\;\; \left. - \int_0^{1-(1-x_1)(1-x_2)}\frac{\dx\zeta}{1-\zeta} \left(\left[\zeta - \iz\, \sgnsi\right]^{-\eps-1} -1 \right) \right\rbrace . \label{eq:D0_regularized_integrals}
	\end{align}
Note that we have rewritten the Gamma functions in the prefactor such that they become unity in the limit $\eps\rightarrow 0$, using ${\Gamma(z+1)=z\,\Gamma(z)}$.

Since the result for the internal-massless single off-shell scalar box integral is commonly given in terms of Gauss hypergeometric functions, we will identify them here, even though this will introduce spurious additional branch cuts. The three integrals appearing in eq.\,\eqref{eq:D0_regularized_integrals} are of the form
	\begin{equation}
		I(\chi) \,\equiv\, \int_0^{\chi}\frac{\dx\zeta}{1-\zeta} \left(\left[\zeta - \iz\,\sgnsi\right]^{-\eps-1}-1\right) .
	\end{equation}
This is trivially zero if $\chi = 0$. All three integrals have vanishing $\chi$ for $s_3 =0$. As mentioned above, this special case of vanishing Mandelstam variables will be considered separately in appendix \ref{app:Vansishing_Mandelstam_variables}. The upper boundary of the third integral,
	\begin{equation}
		1-(1-x_1)(1-x_2) \,=\, -\frac{s_3 q^2}{s_1s_2} \,, \label{eq:third_integral_qsquared_zero}
	\end{equation}
also vanishes if $q^2=0$. We will account for this by setting the third integral to zero in case $q^2=0$. In case $\chi \neq 0$, we can substitute $\zeta \rightarrow \chi^{-1}\zeta$,
	\begin{align}
		I(\chi) \,=\, \int_0^1\frac{\dx\zeta}{1-\chi\zeta} \left(\left[\chi -\iz\,\sgnsi\right]^{-\eps}\zeta^{-\eps-1} - \chi\right) .
		\label{eq:Introducing_iztilde}
	\end{align}
As $\zeta>0$, we could factor it out of the square brackets here. To split this integral into two separate integrals, we must regularize the divergence that occurs for $\zeta \,=\, 1/\chi$ if $\chi > 1$. We regulate it by introducing an infinitesimal imaginary part $\chi \rightarrow \chi + \iztilde$ in the denominator, with $\tilde{0}$ taken to be positive.\footnote{Here, we write $+\iztilde$ as an abbreviation for $\iu\tilde{\eta}$ and taking the limit $\tilde{\eta} \rightarrow +0$. This slightly unusual notation is chosen in analogy to the causal $+\iz$ originating from the propagators, where $+\iz$ is a common short hand notation for $\iu\eta$ and taking the limit $\eta \rightarrow 0$. The tilde marks the new regulator as an analogous but independent limiting process. The distinction between $\iz$ and $\iztilde$ allows us to explicitly check that the end result does not depend on the regulator introduced to split $I(\chi)$ into two integrals.}
Note that the end result will not  depend on the regulator $\iztilde$, since the original integral $I(\chi)$ was not divergent for $\zeta \,=\, 1/\chi$.
Proceeding as described, we obtain
	\begin{align}
		I(\chi) \,&= \left[\chi -\iz\, \sgnsi\right]^{-\eps} \int_0^1\dx\zeta\, \zeta^{-\eps-1} \left(1-(\chi+\iztilde)\zeta\right)^{-1} -\, \chi\int_0^1\dx\zeta\, \frac{1}{1-(\chi + \iztilde)\zeta} \nonumber
		\\
		&=\, -\frac{1}{\eps} \left[\chi -\iz\, \sgnsi\right]^{-\eps} \hyg{1,-\eps,1-\eps;\chi+\iztilde} +\, \ln(1-\chi-\iztilde) \,. \label{eq:I_chi}
	\end{align}
Here, the first integral became a representation of the Gauss hypergeometric function discussed in appendix \ref{app:2F1} (see eq.\,\eqref{eq:IntegralRepresentation_2F1}), while the second integral could be elementarily evaluated in terms of a logarithm. Both hypergeometric function and logarithm are evaluated on the branch cut for $\chi > 1$. The regulator $\iztilde$ tells us which side of these branch cuts to choose. Also note that the prefactor of the hypergeometric function vanishes in case $\chi = 0$, since we assumed $\eps <0$ for $D_0$ to converge.

\begin{figure}[tbp]
		\centering
		\includegraphics[width=0.8\textwidth]{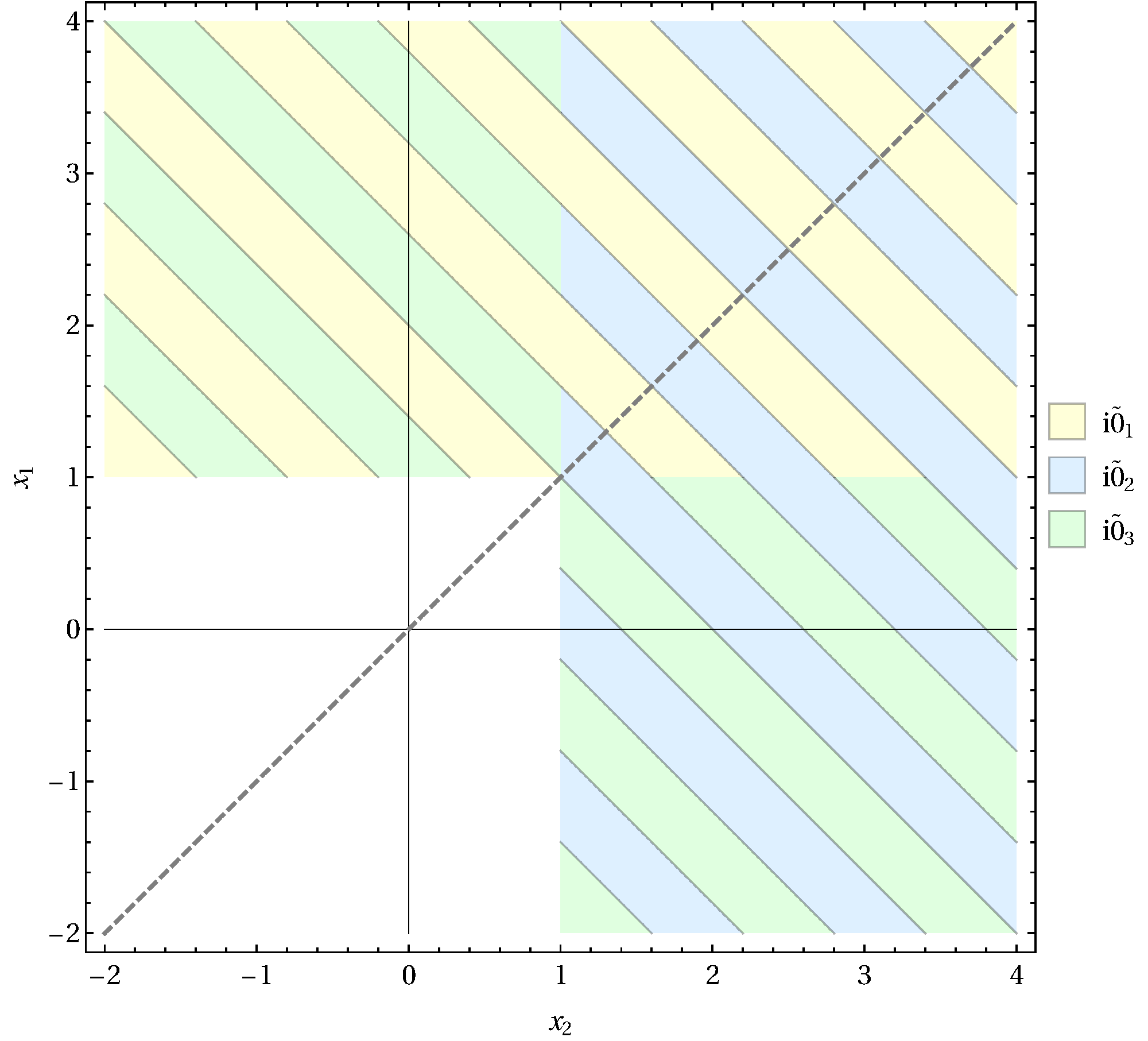}
		\caption{This plot shows in which kinematic regions the respective imaginary parts $\iztilde_i$ of the three logarithms in eq.\,\eqref{eq:imaginary_parts_logs} yield a contribution.}\label{fig:imaginary_parts_logs}
	\end{figure}

Applying eq.\,\eqref{eq:I_chi} to the term in curly braces in eq.\,\eqref{eq:D0_regularized_integrals} and using a different regulator $\iztilde_i$ for each of the three integrals gives 
	\begin{align}
		&-\, \frac{1}{\eps} \left[x_1 -\iz\,\sgnsi\right]^{-\eps} \hyg{1,-\eps,1-\eps;x_1+\iztilde_1} \nonumber
		\\
		&-\, \frac{1}{\eps} \left[x_2 -\iz\,\sgnsi\right]^{-\eps} \hyg{1,-\eps,1-\eps;x_2+\iztilde_2} \nonumber
		\\
		&+\,\frac{1}{\eps} \left[1-(1-x_1)(1-x_2) -\iz\,\sgnsi\right]^{-\eps} \hyg{1,-\eps,1-\eps; 1-(1-x_1)(1-x_2) +\iztilde_3} \nonumber
		\\
		&+\, \ln(1-x_1-\iztilde_1) \,+\, \ln(1-x_2-\iztilde_2) \,-\, \ln\!\left((1-x_1)(1-x_2)-\iztilde_3\right) .
	\end{align}
Writing the logarithms on the last line of this expression in terms of their real and imaginary parts, we see that only the imaginary parts yield a contribution,
	\begin{align}
		&\ln(1-x_1-\iztilde_1) \,+\, \ln(1-x_2-\iztilde_2) \,-\, \ln\!\left((1-x_1)(1-x_2)-\iztilde_3\right) \nonumber
		\\
		& =\, \iu\pi \left[-\, \sgn{\tilde{0}_1} \Theta(x_1-1) \,-\, \sgn{\tilde{0}_2}\, \Theta(x_2-1) \right. \nonumber
		\\
		&\left. \hphantom{=\, \iu\pi} \,+\,\sgn{\tilde{0}_3} \left\lbrace \Theta(x_1-1)\,\Theta(1-x_2) \,+\, \Theta(1-x_1)\,\Theta(x_2-1)\right\rbrace \right] . \label{eq:imaginary_parts_logs}
	\end{align}
Here, $\Theta(x)$ is the Heaviside step function. Let us now choose the sign of the regulators $\iztilde_i$ such that the contribution of the logarithms completely cancels. The kinematic regions in the $x_1x_2$-plane in which the respective regulators yield a contribution are depicted in figure \ref{fig:imaginary_parts_logs}.
For the right-hand side of eq.\,\eqref{eq:imaginary_parts_logs} to vanish, the regulators must satisfy
	\begin{align}
		\sgn{\tilde{0}_1} \,&\stackrel{!}{=}\, \sgn{\tilde{0}_3} \quad \text{if } x_1>1 \text{ and } x_2 <1 \text{ (yellow-green region)}\,,
		\\
		\sgn{\tilde{0}_1} \,&\stackrel{!}{=}\, -\,\sgn{\tilde{0}_2} \quad \text{if } x_1>1 \text{ and } x_2 >1 \text{ (yellow-blue region)}\,,
		\\
		\sgn{\tilde{0}_2} \,&\stackrel{!}{=}\, \sgn{\tilde{0}_3} \quad \text{if } x_1<1 \text{ and } x_2>1 \text{ (blue-green region)}\,.
	\end{align}
In case  both $x_{1,2}<1$ (white region), the right-hand side of eq.\,\eqref{eq:imaginary_parts_logs} vanishes independently of the choice of regulators. Note that since these conditions only determine the relative signs of the regulators, we can choose the absolute sign of one regulator. Let us select ${\iztilde_3 \equiv +\iztilde}$ everywhere. Then, both $\iztilde_1$ and $\iztilde_2$ must change their sign in the yellow-blue region in order to satisfy the conditions. To preserve symmetry, we let them change signs along the dashed diagonal depicted in figure \ref{fig:imaginary_parts_logs}, such that
	\begin{align}
		\iztilde_1 \,&\equiv\, \iztilde\, \sgn{x_1 - x_2} ,
		\\
		\iztilde_2 \,&\equiv\, \iztilde\, \sgn{x_2 - x_1} ,
		\\
		\iztilde_3 \,&\equiv\, \iztilde\,.
	\end{align}
Combining our results, we find that the internal-massless single off-shell scalar box integral is given by
		\begin{align}
			D_0\!\left(s_1,s_2,q^2\right) &=\, \frac{1}{\eps^2} \frac{\Gamma(1+\eps)\, \Gamma^2(1-\eps)}{\Gamma(1-2\eps)}\, \frac{2}{s_1s_2} \left(\frac{s_3\mu^2}{s_1s_2} + \iz\right)^{\!\eps} \nonumber
			\\
			&\times \left\lbrace \left[-\frac{s_3}{s_1} -\iz\,\sgn{\frac{s_3}{s_1s_2}}\right]^{-\eps} \hyg{1,-\eps,1-\eps;\,-\frac{s_3}{s_1}+\iztilde\,\sgn{\frac{s_3}{s_2}-\frac{s_3}{s_1}}} \right. \nonumber
			\\
			&\left. \hphantom{\times\lbrace} \,+ \left[-\frac{s_3}{s_2} -\iz\,\sgn{\frac{s_3}{s_1s_2}}\right]^{-\eps} \hyg{1,-\eps,1-\eps;\, -\frac{s_3}{s_2}+\iztilde\,\sgn{\frac{s_3}{s_1} - \frac{s_3}{s_2}}} \right. \nonumber 
			\\
			&\left. \hphantom{\times\lbrace} \,- \left[-\frac{s_3q^2}{s_1s_2} -\iz\,\sgn{\frac{s_3}{s_1s_2}}\right]^{-\eps} \hyg{1,-\eps,1-\eps;\, -\frac{s_3q^2}{s_1s_2} +\iztilde} \right\rbrace . \label{eq:D0_general_result_two_variable_function}
		\end{align}
Remember that the term in curly braces is a function of the two variables $x_1$ and $x_2$ defined in eq.\,\eqref{eq:Def_x1x2}. The global prefactor $(\ldots)^{\eps}$ and the prefactors $[\ldots]^{-\eps}$ of the hypergeometric functions can be recombined by first inverting the factor in square brackets using eq.\,\eqref{eq:identity_for_inverting} and subsequently applying eq.\,\eqref{eq:identity_for_factoring} from right to left, resulting in
\begin{tcolorbox}[colback=green!10!white]
		\begin{align}
			D_0\!&\left(s_1,s_2,q^2\right) =\, \frac{1}{\eps^2} \frac{\Gamma(1+\eps)\, \Gamma^2(1-\eps)}{\Gamma(1-2\eps)}\, \frac{2}{s_1s_2} \nonumber
			\\
			&\times \left\lbrace \left[\frac{\mu^2}{-s_2-\iz}\right]^{\eps} \hyg{1,-\eps,1-\eps;\,-\frac{s_3}{s_1}+\iztilde\,\sgn{\frac{s_3}{s_2}-\frac{s_3}{s_1}}} \right. \nonumber
			\\
			&\left. \hphantom{\times\lbrace} \,+ \left[\frac{\mu^2}{-s_1-\iz}\right]^{\eps} \hyg{1,-\eps,1-\eps;\, -\frac{s_3}{s_2}+\iztilde\,\sgn{\frac{s_3}{s_1} - \frac{s_3}{s_2}}} \right. \nonumber 
			\\
			&\left. \hphantom{\times\lbrace} \,- \left[\frac{\mu^2}{-q^2-\iz}\right]^{\eps} \hyg{1,-\eps,1-\eps;\, -\frac{s_3q^2}{s_1s_2} +\iztilde} \right\rbrace . \label{eq:D0_general_result_combined_factors}
		\end{align}
\end{tcolorbox}
\noindent{}For convergence, we required $\eps<0$ during the calculation. For the expansion about the physical four dimensions, the result is analytically continued to larger $\eps$.

The result was calculated for ${s_1, s_2, s_3 \in \mathbb{R}\!\setminus\! \lbrace 0\rbrace}$. In appendix \ref{app:Vansishing_Mandelstam_variables}, we will see that it is also valid for vanishing Mandelstam variables in the sense of appropriate limits.
As discussed underneath eq.\,\eqref{eq:third_integral_qsquared_zero}, the term on the last line is equal to zero for ${q^2=0}$, since $\eps<0$. Remember that in dimensional regularization kinematic limits need to be taken before analytic continuation in $\eps$ to assure consistent results.

For the kinematic variables, the causal $\iz$ and the regulator $\iztilde$ specify on which side of the occurring branch cuts the respective functions are evaluated. Remember that by construction this result does not depend on the regulator $\iztilde$. Hence, the individually discontinuous imaginary parts of the hypergeometric functions must cancel each other, and the branch cuts of the hypergeometric functions are spurious. We will see that this is indeed the case in eq.\,\eqref{eq:ExplicitCancellationi0tilde} below.

In the region where $s_1,s_2,q^2<0$ and all three hypergeometric functions are away from their branch cut, our result agrees with those found in \citep[eq.\,(D.4)]{Matsuura1989} and \citep[eq.\,(B.10)]{FabiValery}. The result in \citep[eq.\,(4.36)]{BernDixonKosower1} agrees upon application of \citep[eq.\,(15.3.4)]{Abramovitz_Stegun} to the hypergeometric functions.

Expanding the hypergeometric functions in $\eps$ gives a result in terms of polylogarithms (see eq.\,\eqref{eq:2F1_eps_expansion}).
As long as one avoids the branch cuts of the polylogarithms, the analytic continuation works via $s_1\rightarrow s_1+\iz$, $s_2\rightarrow s_2+\iz$, $q^2\rightarrow q^2+\iz$ as given in \citep{BernDixonKosower1}. However, treating the branch cuts of the polylogarithms by this prescription leads to inconsistent results. Remarkably, the prescription works correctly for the result expanded up to $\mathcal{O}(\eps^0)$ if the Abel identity \citep{Lewin_Polylogs_Associated_Functions} is applied to the dilogarithms first.
Also the analytic continuation procedure introduced in \citep{Beenakker1990} and put forward in \citep{Ellis} only applies to the dilogarithm as it is based on the Euler identity \citep{Lewin_Polylogs_Associated_Functions}.
It is worth mentioning that all these prescriptions for analytic continuation make use of the causal $+\iz$ and apply to each term individually. However from eq{.}\,\eqref{eq:D0_general_result_combined_factors} it is clear that the imaginary parts of the hypergeometric functions respectively polylogarithms are determined by a distinct regulator $\iztilde$ and are interdependent.

\section{Epsilon expansion of the scalar box integral}
\label{sec:Epsilon expansion}
The scalar box integral given in eq.\,\eqref{eq:D0_general_result_two_variable_function} can be expressed as
	\begin{align}
		D_0\!\left(s_1,s_2,q^2\right) &=\, \frac{1}{\eps^2} \frac{\Gamma(1+\eps)\, \Gamma^2(1-\eps)}{\Gamma(1-2\eps)}\, \frac{2}{s_1s_2} \left|\frac{s_3\mu^2}{s_1s_2}\right|^{\!\eps} \nonumber
		\\
		&\hphantom{=}\times\, \exp\!\left[\iu\pi\eps\, \Theta\!\left(-\frac{s_3}{s_1s_2}\right)\right] \mathcal{D}_0\!\left(\eps;-\frac{s_3}{s_1},-\frac{s_3}{s_2}\right) ,
		\label{eq:D0_inTermsOf_CalD0}
	\end{align}
where $\mathcal{D}_0$ abbreviates the term in curly braces. By defining
	\begin{align}
		\mathcal{F}_{\pm}(\eps;x) \,\equiv\, \left(x-\iz\, \sgnsi\right)^{\!-\eps} \hyg{1,-\eps,1-\eps;x\pm\iztilde} ,
		\label{eq:Def_Fpm}
	\end{align}
we can write
	\begin{align}
		\mathcal{D}_0(\eps;x_1,x_2) =\, \Fpm{x_1} \,+\, \Fmp{x_2} \,-\, \Fp{1-(1-x_1)(1-x_2)} \,,
		\label{eq:CalD0_in_terms_of_F}
	\end{align}
where the upper sign choice is valid for $x_1 \geq x_2$ and else the lower sign choice. To expand the scalar box integral around $\eps=0$, we need the $\eps$-expansion of $\Fpm{x}$, which is performed in section \ref{sec:EpsExpansionFpm}. Using this key ingredient, we subsequently discuss the expansion of $ \mathcal{D}_0\!\left(\eps;x_1,x_2\right)$ in section \ref{sec:EpsilonExpansionCalD0} and of $D_0(s_1,s_2,q^2)$ in section \ref{sec:EpsilonExpansionD0}.
 Tabulated results for the real and imaginary parts of $\exp\!\left[\iu\pi\eps\, \Theta\!\left(-\frac{s_3}{s_1s_2}\right)\right] \mathcal{D}_0\!\left(\eps;x_1,x_2\right)$ in the various kinematic regions are given in table \ref{tab: Re and Im of D0} in section \ref{sec:EpsilonExpansionD0}.

\subsection[Epsilon expansion of $\mathcal{F}_{\pm}(\eps;x)$]{Epsilon expansion of \boldmath{$\mathcal{F}_{\pm}(\eps;x)$}} \label{sec:EpsExpansionFpm}
To perform the $\eps$-expansion of $\Fpm{x}$\,, which was defined in eq.\,\eqref{eq:Def_Fpm}, we observe that $\left(x-\iz\,\sgnsi\right)^{-\eps}$ develops an imaginary part for $x<0$, while  the hypergeometric function does so for $x>1$. Hence, there are 3 regions of interest:
	\begin{itemize}
		\item[1.] $x\in(-\infty,0)$, where we need to expand $\left(x-\iz\,\sgnsi\right)^{-\eps}\hyg{1,-\eps,1-\eps,x}$,
		\item[2.] $x\in(0,1)$, where we need to expand  $x^{-\eps}\,\hyg{1,-\eps,1-\eps,x}$,
		\item[3.] $x\in(1,\infty)$, where we need to expand $x^{-\eps}\,\hyg{1,-\eps,1-\eps,x\pm\iztilde}$.
	\end{itemize}

\subsubsection[Expansion for $x\in(-\infty,0)$]{Expansion for \boldmath{$x\in(-\infty,0)$}}
For $x<0$, $\Fpm{x}$ reads
	\begin{align}
		\Fpm{x} \,&= \left(x-\iz\, \mathrm{sgn}_{123}\right)^{-\eps}\hyg{1,-\eps,1-\eps,x} \nonumber
		\\
		&=\, \e^{\iu\pi\eps\, \sgnsi}|x|^{-\eps} \left(1-\sum_{n=1}^\infty\eps^n\,\Li{n}{x}\right) \nonumber
		\\
		&=\, \e^{\iu\pi\eps\, \sgnsi}\left[|x|^{-\eps}-\sum_{k=0}^\infty\frac{(-\eps)^k}{k!}\ln^k|x|\sum_{n=1}^\infty\eps^n\,\Li{n}{x}\right],
	\end{align}
where the $\eps$-expansion of the hypergeometric function given in eq.\,\eqref{eq:2F1_eps_expansion} was used in the first step. Next, we substitute $n \rightarrow N = n+k$ and then interchange the $N$- and $k$-sum,
	\begin{align}
		\Fpm{x} \,&=\, \e^{\iu\pi\eps\, \sgnsi}\left[|x|^{-\eps}-\sum_{N=1}^\infty\eps^N\sum_{k=0}^{N-1}\frac{(-1)^k\ln^k|x|}{k!}\,\Li{N-k}{x}\right].
	\end{align}
For $N\geq 2$, we now introduce the \textit{single-valued polylogarithms} $\LiSV{n}{x}\,$ discussed in appendix \ref{app:SingleValuedPolylogs} by adding and subtracting $\frac{(-1)^{N-1}}{N!}\ln^{N-1}|x|\ln|1-x|$ in the sum. This yields
	\begin{align}
		\Fpm{x} \,&=\, \e^{\iu\pi\eps\, \sgnsi} \left[\vphantom{\frac{(-1)^{N-1}}{N!}}|x|^{-\eps} \,-\, \eps\, \Li{1}{x} \right.\nonumber
		\\
		&\qquad\qquad\quad\left.\,-\, \sum_{N=2}^\infty \eps^N \left(\LiSV{N}{x} \,-\, \frac{(-1)^{N-1}}{N!}\ln^{N-1}|x|\ln|1-x|\right)\right]
		\nonumber
		\\
		&=\e^{\iu\pi\eps\, \sgnsi} \left[1 \,+\, \ln\frac{x}{x-1} \sum_{n=1}^\infty\frac{(-\eps)^{n}}{n!}\ln^{n-1}|x|-\sum_{n=2}^\infty\eps^n\LiSV{n}{x}\right] .
		\label{eq:Fexpansion x<0}
	\end{align}
Note that since the single-valued polylogarithms are bounded on $\mathbb{R}$, any divergences in the kinematic variable $x$ are explicitly contained in the logarithms.

\subsubsection[Expansion for $x\in(0,1)$]{Expansion for \boldmath{$x\in(0,1)$}}
In the area where $0<x<1$, neither $(x-\iz\,\sgnsi)^{-\eps}$ nor $\hyg{1,-\eps,1-\eps;x\pm\iztilde}$ develop an imaginary part. Following the same steps as before, we find
	\begin{align}
		\Fpm{x} \,&=\, x^{-\eps}\, \hyg{1,-\eps,1-\eps,x}
		\nonumber\\
		&=\, |x|^{-\eps} \left(1-\sum_{n=1}^\infty\eps^n\,\Li{n}{x}\right)
		\nonumber\\
		&=\, |x|^{-\eps}  \,-\, \eps\,\Li{1}{x} \,-\, \sum_{n=2}^\infty \eps^n \left(\LiSV{n}{x} \,-\, \frac{(-1)^{n-1}}{n!}\ln^{n-1}|x|\ln|1-x|\right)
		\nonumber\\
		&=1+\ln\frac{x}{1-x} \sum_{n=1}^\infty\frac{(-\eps)^{n}}{n!}\ln^{n-1}|x|-\sum_{n=2}^\infty\eps^n\LiSV{n}{x} .
		\label{eq:Fexpansion 0<x<1}
	\end{align}
Again, any divergences in the kinematic variable are contained in the logarithms since the single-valued polylogarithms are bounded on $\mathbb{R}$.

\subsubsection[Expansion for $x\in(1,\infty)$]{Expansion for \boldmath{$x\in(1,\infty)$}}
In the area where $x>1$, the hypergeometric function $\hyg{1,-\eps,1-\eps,x\pm\iztilde}$ develops an imaginary part. To make it explicit, we employ the inversion formula given in eq.\,\eqref{eq:2F1_inversion_formula},
	\begin{align}
		\Fpm{x} \,&=\, x^{-\eps}\hyg{1,-\eps,1-\eps,x\pm\iztilde}
		\nonumber\\
		&=\, |x|^{-\eps}\left[-\hyg{1,\eps,1+\eps;\frac{1}{x}} \,+\, 1 \,+\, (-x\mp\iztilde)^{\eps} \frac{\pi \eps}{\sin(\pi\eps)}\right].
	\end{align}
By replacing $\eps\rightarrow-\eps$, we can use the $\eps$-expansion from eq.\,\eqref{eq:2F1_eps_expansion} for the resulting hypergeometric function. Additionally writing the last term in terms of its real and imaginary part, we obtain
	\begin{align}
		\Fbox{\pm}{x;\eps} \,&=\, |x|^{-\eps}\, \sum_{n=1}^\infty(-\eps)^n\, \Li{n}{\frac{1}{x}} \,+\, \pi\eps\cot(\pi\eps) \,\mp\, \iu\pi\eps\,.
	\end{align}
The $\eps$-expansion of the cotangent term is given by \citep[eqs.\,(4.3.70) \& (23.1.18)]{Abramovitz_Stegun}
	\begin{align}
		\pi \eps\cot(\pi \eps) \,=\, 1 \,-\, 2\sum_{n=1}^\infty\zeta_{2n}\,\eps^{2n} \,=\, 1-\sum_{n=2}^\infty\,[1+(-1)^n]\,\zeta_n \eps^n\,,
	\end{align}
where $\zeta_n=\zeta(n)=\sum_{k=1}^\infty k^{-n}$ are the integer values of the Riemann zeta function.
Additionally expanding $|x|^{-\eps}$ and collecting orders of $\eps$ yields
	\begin{align}
		\Fpm{x} \,=\, &1 \,-\, \eps\, \Li{1}{\frac{1}{x}} \,\mp\, \iu\pi\eps
		\nonumber\\
		&-\, \sum_{n=2}^\infty\eps^n \left[(-1)^{n-1} \sum_{k=0}^{n-1}\frac{(-1)^k}{k!}\, \ln^k\left|\frac{1}{x}\right|\, \Li{n-k}{\frac{1}{x}}+[1+(-1)^n]\zeta_n\right].
	\end{align}
Upon adding and subtracting the term $\frac{(-1)^{n-1}}{n!}\ln^{n-1}\left|\frac{1}{x}\right|\ln\left|1-\frac{1}{x}\right|$, we can identify the single-valued polylogarithms $\LiSV{n}{\frac{1}{x}}$, as defined in eq{.}\,\eqref{eq:Definiton SVP}. Inverting them with the help of eq.\,\eqref{eq:PolyLog inversion relation}, we find
	\begin{align}
		\Fpm{x} \,=\, 1 \,-\, \eps\,\Li{n}{\frac{1}{x}} \,\mp\, \iu\pi\eps \,-\, \sum_{n=2}^\infty \eps^n \left[\LiSV{n}{x} -\, \frac{(-1)^{n-1}}{n!}\ln^{n-1}\left|x\right|\ln\left|\frac{x-1}{x}\right|\right] .
	\end{align}
Organizing this expression in terms of logarithms and single-valued polylogarithms yields
	\begin{align}
		\Fpm{x} \,=\, 1 \,+\, \ln\frac{x}{x-1} \sum_{n=1}^\infty \frac{(-\eps)^n}{n!}\ln^{n-1}|x| \,-\, \sum_{n=2}^\infty\eps^n\LiSV{n}{x} \,\mp\, \iu\pi\eps \,.
		\label{eq:Fexpansion x>1}
	\end{align}

\subsubsection[Representation of the expansion of $\Fpm{x}$ on the entire real axis]{Representation of the expansion of \boldmath{$\Fpm{x}$} on the entire real axis}
\label{sec:Fcal epsilon expansion on the real axis}
Combining the results found in eqs.\,\eqref{eq:Fexpansion x<0}, \eqref{eq:Fexpansion 0<x<1}, and \eqref{eq:Fexpansion x>1}, we can give a representation for the $\eps$-expansion of $\Fpm{x}$ valid for the entire domain of $x\in\mathbb{R}$,
	\begin{tcolorbox}[colback=green!10!white]
		\begin{align}
			&\Fpm{x} \,=\, \e^{\iu\pi\eps\, \sgnsi\, \Theta(-x)} \mathfrak{F}(\eps;x) \,\mp\, \iu\pi\eps\, \Theta(x-1)\,,
			\label{eq:Fexpansion result}
			\\
			\text{where} \quad &\mathfrak{F}(\eps;x) \,\equiv\, 1 \,+\, \ln\!\left|\frac{x}{x-1}\right| \sum_{n=1}^\infty \frac{(-\eps)^n}{n!} \ln^{n-1}|x| \,-\, \sum_{n=2}^\infty\eps^n\LiSV{n}{x} \,.
			\label{eq:frakFexpansion result}
		\end{align}
	\end{tcolorbox}
\noindent{}This compact expression is well suited to be used for the $\eps$-expansion of the complete box integral. The imaginary part is made explicit and all functions of the real variable $x$ are manifestly real. The complex exponential stems from $\left(x-\iz\,\sgnsi\right)^{-\eps}$, while the imaginary part $\pi\eps\,\Theta(x-1)$ originates from the hypergeometric function. %$\hyg{1,-\eps,1-\eps;x\pm\iztilde}$

An important feature of the expansion of $\mathfrak{F}$ as given in eq.\,\eqref{eq:frakFexpansion result} is that it is finite in the limit $x\rightarrow\pm\infty$ in every order in $\eps$. Using the limits of the single-valued polylogarithms given in appendix \ref{app:SingleValuedPolylogs}, we obtain
	\begin{align}
		\lim_{x\rightarrow\infty}\mathfrak{F}(\eps;x) \,&=\, 1 \,-\, 2\sum_{n=1}^\infty \eps^{2n}\zeta_{2n}\,,
		\label{eq:Ffrak limiting behavior +infty}
	   \\
		\lim_{x\rightarrow-\infty}\mathfrak{F}(\eps;x) \,&=1\,-\, 2\sum_{n=1}^\infty \eps^{2n}(2^{1-2n}-1)\zeta_{2n}\,.
		\label{eq:Ffrak limiting behavior -infty}
	\end{align}
By collecting the $\ln|x|$ terms we can cast eq{.}\,\eqref{eq:frakFexpansion result} into the alternative form
\begin{align}
\mathfrak{F}(\eps;x)=|x|^{-\eps}+\eps\ln|1-x|-\sum_{n=2}^\infty \eps^n \left[\frac{(-1)^n\ln|1-x|\ln^{n-1}|x|}{n!}+\LiSV{n}{x}\right].
\label{eq:Ffrak limiting behavior 0}
\end{align}
From this we can read off that $\mathfrak{F}(\eps;x)$ behaves like $|x|^{-\eps}$ near $x=0$ and it diverges logarithmically like $\eps\ln|1-x|$ near $x=1$ . In the complete box integral the $\ln|1-x|$ divergences always cancels.

\subsection[Epsilon expansion of $\mathcal{D}_0(\eps;x_1,x_2)$]{Epsilon expansion of \boldmath{$\mathcal{D}_0(\eps;x_1,x_2)$}} \label{sec:EpsilonExpansionCalD0}
	\begin{figure}[bth]
		\centering
		\includegraphics[width=0.8\textwidth]{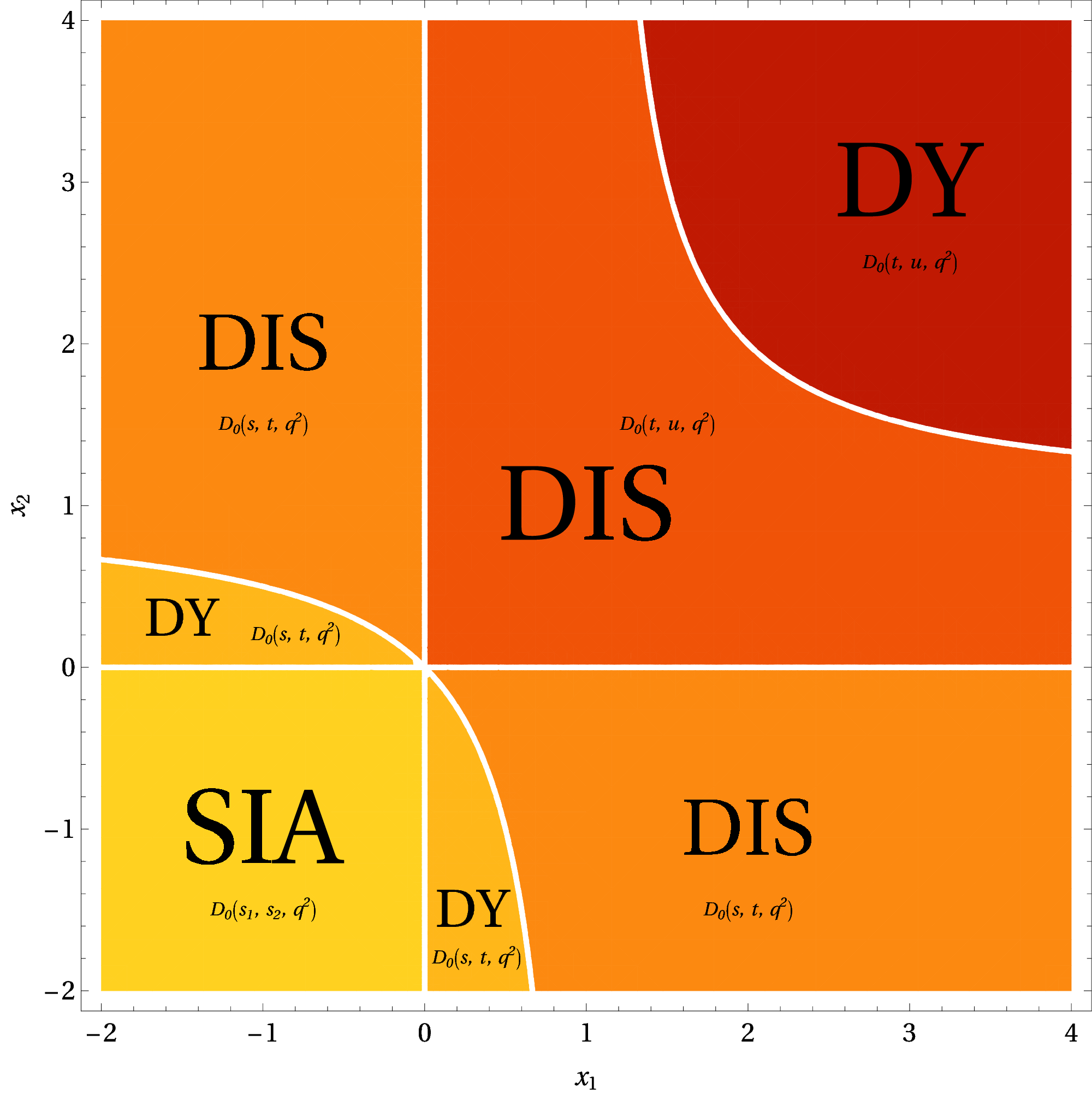}
		\caption{Physical interpretation of the kinematic regions to be considered for the $\eps$-expansion of $\mathcal{D}_0(\eps;x_1,x_2)$. For DIS and Drell-Yan, the Mandelstam variables are $s>0$, $t<0$, and $u<0$. In the case of SIA, all Mandelstam variables are positive. The virtuality $q^2$ of the off-shell particle is negative for DIS and positive for both Drell-Yan and SIA}
		\label{fig:x1x2planeDISDYHP}
	\end{figure} \noindent{}
By plugging the $\eps$-expansion of $\Fpm{x}$ from eq{.}\,\eqref{eq:Fexpansion result} into eq{.}\,\eqref{eq:CalD0_in_terms_of_F}, it is straightforward to obtain the $\eps$-expansion of $\mathcal{D}_0(\eps;x_1,x_2)$. We must consider seven areas corresponding to different kinematics, which are depicted in figure \ref{fig:x1x2planeDISDYHP}. Since $\mathcal{D}_0(\eps;x_1,x_2)$ is symmetric under exchanging $x_1 \leftrightarrow x_2$, only five of these areas are distinct. There are
	\begin{itemize}
		\item[1.] DIS kinematics with $x_1$ and $x_2$ both positive,
		\item[2.] DIS kinematics where $x_1$ and $x_2$ have different signs,
		\item[3.] Drell-Yan kinematics with $x_1$ and $x_2$ both positive,
		\item[4.] Drell-Yan kinematics where $x_1$ and $x_2$ have different signs,
		\item[5.] SIA kinematics with $x_1$ and $x_2$ both negative.
	\end{itemize}

Observing that
	\begin{align}
		\iu\pi\eps &\left[-\,\sgn{x_1-x_2}\,\Theta(x_1-1) \,-\,\sgn{x_2-x_1}\,\Theta(x_2-1) \,+\, \Theta(-(1-x_1)(1-x_2)) \right] =\, 0 \,,
		\label{eq:ExplicitCancellationi0tilde}
	\end{align}
we see that the imaginary parts originating from the hypergeometric functions cancel in all kinematic regions, as they should by construction. Hence, we obtain
	\begin{align}
		\mathcal{D}_0(\eps;x_1,x_2) \,=\, &\e^{\iu\pi\eps\,\sgnsi\Theta(-x_1)}\mathfrak{F}(\eps;x_1) \,+\, \e^{\iu\pi\eps\,\sgnsi\Theta(-x_2)}\mathfrak{F}(\eps;x_2)
		\nonumber\\
		&-\, \e^{\iu\pi\eps\,\sgnsi\Theta(-x_1-x_2+x_1 x_2)}\mathfrak{F}(\eps;x_1+x_2-x_1 x_2) \,.
		\label{eq: Dcal0 with Ffrak}
	\end{align}

The $\eps$-expansion of $\mathcal{D}_0$ in the first region, where none of the three parts develops an imaginary part, is purely real. The logarithmic poles at $x_{1,2}=1$ cancel between the three $\mathfrak{F}$-functions, such that
\begin{tcolorbox}[colback=green!10!white]
	\begin{align}
		&\mathcal{D}_0(\eps;x_1,x_2) \,=\, \mathfrak{F}\!\left(\eps;x_1\right) \,+\, \mathfrak{F}\!\left(\eps;x_2\right) \,-\, \mathfrak{F}\!\left(\eps;x_1+x_2-x_1x_2\right)
		\nonumber\\
		&=\, 1 \,-\, \eps\ln\left|\frac{x_1 x_2}{x_1+x_2-x_1 x_2}\right|+\sum_{n=2}^\infty\frac{(-\eps)^n}{n!}\left[\ln\left|\frac{x_1}{x_1-1}\right|\ln^{n-1}|x_1|\right.
		\nonumber\\
		&\phantom{=}\left. +\ln\left|\frac{x_2}{x_2-1}\right| \ln^{n-1}|x_2|-\ln\left|\frac{x_1+x_2-x_1x_2}{(1-x_1)(1-x_2)}\right| \ln^{n-1}|x_1+x_2-x_1x_2|\right]
		\nonumber\\
		&\phantom{=}-\, \sum_{n=2}^\infty \eps^n \left[\LiSV{n}{x_1}+\LiSV{n}{x_2}-\LiSV{n}{x_1+x_2-x_1x_2}\right] .
		\label{eq:CalD0_EpsExpansion_x1x2positive}
	\end{align}
\end{tcolorbox}\noindent{}
%To refer to this purely real epsilon expansion, we define
%\begin{align}
%	\mathcal{D}^\text{eucl.}_0(\eps;x_1,x_2) \,\equiv\, \mathcal{D}_0(\eps;x_1>0,x_2>0) \,=\, \text{eq.}\,\eqref{eq:CalD0_EpsExpansion_x1x2positive} \,,
%\end{align}
%since the same purely real result would be obtained if all Mandelstam variables were space-like euclidean vectors (see eq.\,\eqref{eq:D0_FeynmanParametrized}). 
For the order $\eps^2$ coefficient, we can use Abel's formula \eqref{eq:Abel Li2}, in combination with the Euler identity \eqref{eq:Euler Li2} and inverted Landen identity \eqref{eq:Inverted Landen Li2}, to reduce the number of appearing $\LinSV{2}$ to just two,
\begin{align}
	\LiSV{2}{x_1}+\LiSV{2}{x_2}-\LiSV{2}{x_1+x_2-x_1x_2}=\LiSV{2}{\frac{x_2(x_1-1)}{x_1}}+\LiSV{2}{\frac{x_1(x_2-1)}{x_2}}+\zeta_2\,,
\end{align}
where we assumed $x_1,\, x_2,\, (x_1+x_2-x_1 x_2)>0$.
For the higher orders in $\eps$ there presumably exists no representation with fewer than three (single-valued) polylogarithms, since only for the dilogarithm there exists a suitable five term relation (see appendix \ref{sec:Functional equations of continuous single-valued polylogarithms}).

\subsection[Epsilon expansion of $D_0\!\left(s_1,s_2,q^2\right)$]{Epsilon expansion of \boldmath{$D_0\!\left(s_1,s_2,q^2\right)$}}
\label{sec:EpsilonExpansionD0}
 Combining eqs{.}\,\eqref{eq:D0_inTermsOf_CalD0} and \eqref{eq: Dcal0 with Ffrak}, we see that real and imaginary part of $D_0(s_1,s_2,q^2)$ are determined by the phase factors $\exp\left[{\iu\pi\eps\,\Theta\!\left(-\frac{s_3}{s_1 s_2}\right)}\right]\!\cdot\exp\left[{\iu\pi\eps\,\sgnsi \Theta(-y_i)}\right]$, where ${y_i=-\frac{s_3 s_i}{s_1 s_2}}$ with $s_i=s_1,s_2,q^2$. 
 We can simplify this factor to
 \begin{align}
 	\e^{\iu\pi\eps\Theta\left(-\frac{s_3}{s_1 s_2}\right)}\cdot\e^{\iu\pi\eps\,\sgnsi \Theta(-y_i)}=\Theta(-s_i)+\Theta(s_i)\,\e^{\iu \pi\eps}\,.
 \end{align}
Therefore, $\e^{\iu\pi\eps\Theta\left(-\frac{s_3}{s_1 s_2}\right)}\mathcal{D}_0(\eps,x_1,x_2)$ admits for the representation
	\begin{align}
		\e^{\iu\pi\eps\Theta\left(-\frac{s_3}{s_1 s_2}\right)} \mathcal{D}_0(\eps,x_1,x_2)
		\,=& \left(\Theta(-s_2) + \Theta(s_2)\, \e^{\iu \pi\eps}\right) \mathfrak{F}(\eps;x_1)
		\nonumber\\
		&+ \left(\Theta(-s_1)+\Theta(s_1)\,\e^{\iu\pi\eps}\right) \mathfrak{F}(\eps;x_2)
		\nonumber\\
		&-\left(\Theta(-q^2)+\Theta(q^2)\,\e^{\iu \pi\eps}\right) \mathfrak{F}(\eps;x_1+x_2-x_1 x_2)\,.
	\end{align}
Hence, the imaginary part is determined by the signs of $s_1$, $s_2$ and $q^2$. Explicit results for the real and imaginary part in all areas are given in table \ref{tab: Re and Im of D0}.
	\begin{table}
	\centering
	\begin{tabular}{ccccc}
		\hline
		$s_1$ & $s_2$ & $q^2$ & $\mathrm{Re}\!\left(\e^{\iu\pi\eps\Theta\left(-\frac{s_3}{s_1 s_2}\right)}\mathcal{D}_0(\eps,x_1,x_2)\right)$ & $\mathrm{Im}\!\left(\e^{\iu\pi\eps\Theta\left(-\frac{s_3}{s_1 s_2}\right)}\mathcal{D}_0(\eps,x_1,x_2)\right)$
		\\
		\hline
		$-$ & $-$ & $-$ & $\mathfrak{F}_1+\mathfrak{F}_2-\mathfrak{F}_{12}$ & 0
		\\
		$-$ & $-$ & + & $\mathfrak{F}_1+\mathfrak{F}_2-\mathfrak{F}_{12} \cos\pi \eps$ & $-\mathfrak{F}_{12}\sin\pi\eps$ 
		\\
		$-$ & + & $-$ & $\mathfrak{F}_2-\mathfrak{F}_{12}+\mathfrak{F}_1\cos\pi\eps$ & $\mathfrak{F}_1\sin\pi\eps$
		\\
		+ & $-$ & $-$ & $\mathfrak{F}_1-\mathfrak{F}_{12}+ \mathfrak{F}_2\cos\pi\eps$ & $\mathfrak{F}_2\sin\pi\eps$
		\\
		$-$ & + & + & $\mathfrak{F}_2+(\mathfrak{F}_1-\mathfrak{F}_{12})\cos\pi\eps$ & $(\mathfrak{F}_1-\mathfrak{F}_{12}) \sin\pi\eps$
		\\
		+ & $-$ & + & $\mathfrak{F}_1+(\mathfrak{F}_2-\mathfrak{F}_{12})\cos\pi\eps$ & $(\mathfrak{F}_2-\mathfrak{F}_{12})\sin\pi\eps$
		\\
		+ & + & $-$ & $-\mathfrak{F}_{12}+(\mathfrak{F}_1+\mathfrak{F}_{2})\cos\pi\eps$ & $(\mathfrak{F}_1+\mathfrak{F}_{2})\sin\pi\eps$
		\\
		+ & + & + & $\left(\mathfrak{F}_1+\mathfrak{F}_2-\mathfrak{F}_{12}\right) \cos\pi\eps$ & $\left(\mathfrak{F}_1+\mathfrak{F}_2-\mathfrak{F}_{12}\right)\sin\pi\eps$
		\\
		\hline
		\end{tabular}
	\caption{Real and imaginary part of $\e^{\iu\pi\eps\Theta\left(-\frac{s_3}{s_1 s_2}\right)}\mathcal{D}_0(\eps,x_1,x_2)$ in the areas specified by the signs of $s_1$, $s_2$ and $q^2$. For compactness we introduce the abbreviations $\mathfrak{F}_1\equiv\mathfrak{F}(\eps;x_1)$, $\mathfrak{F}_2\equiv\mathfrak{F}(\eps;x_2)$ and $\mathfrak{F}_{12}\equiv\mathfrak{F}(\eps;x_1+x_2-x_1 x_2)$.}
	\label{tab: Re and Im of D0}
	\end{table}
Note that the superficially divergent terms $\ln(1-x_1)$ and $\ln(1-x_2)$ cancel between $\mathfrak{F}(\eps;x_{1,2})$ and $\mathfrak{F}(\eps;x_1+x_2-x_1 x_2)$ in the regions where $x_{1,2}=1$.

Combining the results listed in table \ref{tab: Re and Im of D0} and eq{.}\,\eqref{eq:D0_inTermsOf_CalD0} we can directly infer real and imaginary parts of the complete box integral $D_0(s_1,s_2,q^2)$.
For completing the $\eps$-expansion we can employ \citep[eq.\,(6.1.33)]{Abramovitz_Stegun}
	\begin{align}
		\Gamma(1+\eps)=\exp\!\left[-\gamma_\text{E}\,\eps+\sum_{n=2}^\infty\frac{(-1)^n \zeta_n\,\eps^n}{n}\right]
	\end{align}
for all Gamma functions in the prefactor, where $\gamma_\text{E}$ is the Euler-Mascheroni constant. However, since the same factor of Gamma functions multiplies all one-loop integrals and a similar factor is present in corresponding phase-space integrals it is often not required to expand all of them explicitly.
Expanding the term $\left|\frac{s_3 \mu^2}{s_1 s_2}\right|^\eps$ introduces additional logarithmic terms in the $\eps$-expansion. One might albeit need to keep this term unexpanded to regularize singular behavior in the kinematic limits $s_1,s_2,s_3\rightarrow 0$.
This is exemplified in appendix \ref{app:Regularizing_kinematic_limits}.

To sum up our result, we can cast $D_0(s_1,s_2,q^2)$ as given in eq{.}\,\eqref{eq:D0_general_result_two_variable_function} into the form
\begin{tcolorbox}[colback=green!10!white]
\begin{align}
		D_0\!\left(s_1,s_2,q^2\right) =\, &\frac{1}{\eps^2} \frac{\Gamma(1+\eps)\, \Gamma^2(1-\eps)}{\Gamma(1-2\eps)}\, \frac{2}{s_1s_2} \left|\frac{s_3\mu^2}{s_1s_2}\right|^{\!\eps} \nonumber
		\\
		\times\,& \left[(\Theta(-s_2)+\Theta(s_2)\e^{\iu\pi\eps})\,\mathfrak{F}(\eps;x_1)
		+(\Theta(-s_1)+\Theta(s_1)\e^{\iu\pi\eps})\,\mathfrak{F}(\eps;x_2)\right.
		\nonumber\\
		&\left.-(\Theta(-q^2)+\Theta(q^2)\e^{\iu\pi\eps})\,\mathfrak{F}(\eps;x_1+x_2-x_1 x_2)\right].
		\label{eq:D0 result with theta functions}
\end{align}
\end{tcolorbox}
\noindent{}Plugging in the $\eps$-expansion of $\mathfrak{F}(\eps;x)$ from eq{.}\,\eqref{eq:frakFexpansion result} we have an all-order $\eps$-expansion of $D_0(s_1,s_2,q^2)$ with explicit real and imaginary parts valid in the entire kinematic domain of $s_1,s_2,q^2$. Since the term in square brackets is finite for vanishing $s_{1,2}$, $D_0(s_1,s_2,q^2)$ behaves like $|s_{1,2}|^{-1-\eps}$ in the limits $s_{1,2}\rightarrow 0$. For a comprehensive discussion of the behavior of $\mathfrak{F}(\eps;x)$ in different kinematic limiting cases we refer to section \ref{sec:Fcal epsilon expansion on the real axis}.

By bringing the factor $\left|\frac{s_3 \mu^2}{s_1 s_2}\right|^\eps$ into the brackets, we obtain a form analogous to eq{.}\,\eqref{eq:D0_general_result_combined_factors},
\begin{tcolorbox}[colback=green!10!white]
\begin{align}
D_0\!\left(s_1,s_2,q^2\right) &=\, \frac{1}{\eps^2} \frac{\Gamma(1+\eps)\, \Gamma^2(1-\eps)}{\Gamma(1-2\eps)}\, \frac{2}{s_1s_2} \nonumber
		\\
		\times&\left[\left(\frac{\mu^2}{-s_2-\iu 0}\right)^\eps\,\left|\frac{s_3}{s_1}\right|^\eps\,\mathfrak{F}\!\left(\eps;-\frac{s_3}{s_1}\right)
		+\left(\frac{\mu^2}{-s_1-\iu 0}\right)^\eps\,\left|\frac{s_3}{s_2}\right|^\eps\,\mathfrak{F}\!\left(\eps;-\frac{s_3}{s_2}\right)\right.
		\nonumber\\
		&\left.-\left(\frac{\mu^2}{-q^2-\iu 0}\right)^\eps\,\left|\frac{s_3 q^2}{s_1 s_2}\right|^\eps\,\mathfrak{F}\!\left(\eps;-\frac{s_3 q^2}{s_1s_2}\right)\right].
\label{eq: D0 in terms of frakF}
\end{align}
\end{tcolorbox}
\noindent{}Eqs{.}\,\eqref{eq:D0_general_result_combined_factors} and \eqref{eq: D0 in terms of frakF} differ only by the replacement $\hyg{1,-\eps,1-\eps;x\pm\iu\tilde{0}} \rightarrow |x|^\eps\,\mathfrak{F}(\eps;x)$.
This can be understood as replacing the hypergeometric function by a suitable single-valued version.
Hence, in contrast to eq.\,\eqref{eq:D0_general_result_combined_factors}, the new eq.\,\eqref{eq: D0 in terms of frakF} does no longer suffer from spurious branch cuts.

\section{Conclusion}
We have revisited the massless scalar box integral with one external off-shell particle, keeping the causal $+\iz$ throughout the calculation. 
The main new result is extending the explicit determination of real and imaginary parts beyond finite order in the dimensional regularization parameter epsilon in all kinematic regions.
The epsilon expansion is expressed to all orders in terms of newly introduced single-valued polylogarithms. This makes our result free of the spurious branch cuts common to representations using ordinary polylogarithms. The single-valued polylogarithms are bounded, such that any divergent behavior in kinematic limits is contained in logarithmic terms.
Since we have determined the box integral in all kinematic regions, our result is suitable for all kinds of calculations in higher orders of perturbation theory, where imaginary parts or higher orders in epsilon become important.

The calculation method put forward in this work can be generalized to the scalar box integral with two non-adjacent end points off the light cone. A corresponding paper is in preparation \citep{Haug2023}.

\acknowledgments
We are grateful to Werner Vogelsang for valuable comments regarding this work. We thank the referee for raising the question if generalizing the result to more end points off the light cone is possible. This study was supported in part by Deutsche
Forschungsgemeinschaft (DFG) through the Research Unit
FOR 2926 (Project No. 409651613). J. H. is grateful to the Landesgraduiertenf{\"o}rderung Baden-W{\"u}rttemberg for supporting her research.
%%%%%%%%%%%%%%%%%%%%%%%%%%%%%%%%%%%%%%%%%%%%%%%%%%%%%%%%%%%%%%%%%%%%%%%%%%%%%%%%%%%%%%%
%
% The appendix begins here
%
%%%%%%%%%%%%%%%%%%%%%%%%%%%%%%%%%%%%%%%%%%%%%%%%%%%%%%%%%%%%%%%%%%%%%%%%%%%%%%%%%%%%%%%
\appendix %\numberwithin{equation}{section}
\section{Scalar box integral in case of vanishing Mandelstam variables}\label{app:Vansishing_Mandelstam_variables}
In this appendix, we evaluate the internal-massless single off-shell scalar box integral in case one or several Mandelstam variables vanish, since the result calculated in the main text was obtained for non-vanishing Mandelstam variables.
For this, we go back to $D_0$ as given in eq.\,\eqref{eq:D0_xi_integrals} and explicitly set the vanishing Mandelstam variables to zero.
We will find that all cases agree with appropriate limits of the general result.

\subsection[$s_1=0$ or $s_2=0$]{\boldmath{$s_1=0$} or \boldmath{$s_2=0$}}
Since the scalar box integral is symmetric under exchanging $s_1\leftrightarrow s_2$, we can without loss of generality consider $s_1=0$. In this case, the integrals in eq.\,\eqref{eq:D0_xi_integrals} factorize,
	\begin{align}
		D_0\!\left(s_1=0, s_2, q^2=s_2+s_3\right) &=\, \mu^{2\eps}\, \frac{\Gamma(2+\eps) \Gamma^2(-\eps)}{\Gamma(-2\eps)} \!\int_0^1\!\!\dx\xi_1 \!\int_0^1\!\!\dx\xi_2\, \left[\xi_2\left(- s_2 - s_3\xi_1 -\iz\right)\right]^{-\eps-2} \nonumber
		\\
		&=\, \frac{2}{\eps} \frac{\Gamma(1+\eps) \Gamma^2(1-\eps)}{\Gamma(1-2\eps)}\, \mu^{2\eps}\!\int_0^1\!\!\dx\xi_1 \left[ -s_2-s_3\xi_1 -\iz\right]^{-\eps-2} .
		\label{eq:D0_s1_vanishing_integral}
	\end{align}
Here the $\xi_2$-integration was performed elementarily. Note that $\eps < -1$ was imposed for convergence. While the $\xi_1$-integral can in principle be evaluated elementarily as well, first substituting ${\xi_1 \rightarrow \zeta = -s_2-s_3\xi_1}$ is useful for discussing the additional limits $q^2,s_{2}=0$. Splitting the resulting integral into two separate integrals, we obtain
	\begin{align}
		D_0\!\left(s_1=0, s_2, q^2=s_2+s_3\right) &=\, \frac{1}{\eps} \frac{\Gamma(1+\eps) \Gamma^2(1-\eps)}{\Gamma(1-2\eps)}\, \frac{2}{s_3}\, \mu^{2\eps} \nonumber
		\\
		&\hphantom{=}\times \left\lbrace\int_{0}^{-s_2}\dx\zeta \left[ \zeta -\iz\right]^{-\eps-2} - \int_0^{-q^2}\dx\zeta \left[ \zeta -\iz\right]^{-\eps-2} \right\rbrace \nonumber
		\\
		&=\, \frac{1}{\eps (1+\eps)} \frac{\Gamma(1+\eps) \Gamma^2(1-\eps)}{\Gamma(1-2\eps)}\, \frac{2}{s_3} \nonumber
		\\
		&\hphantom{=}\times \left\lbrace \frac{1}{s_2}\left(\frac{\mu^2}{-s_2-\iz}\right)^{\!\eps}- \frac{1}{q^2}\left(\frac{\mu^2}{-q^2-\iz}\right)^{\!\eps} \right\rbrace . \label{eq:D0_s1_vanishing}
	\end{align}
Note that the second integral and thus the second term on the last line vanishes for $q^2=s_2+s_3=0$. The result can be analytically continued to values of $\eps$ beyond the original area of convergence.

Let us compare this result to the limit $s_1\rightarrow 0$ of the general result as given eq.\,\eqref{eq:D0_general_result_combined_factors}.
The limit of the first and last term in the general result are of the form
	\begin{equation}
		\lim_{s_1\rightarrow 0}\; \frac{1}{s_1}\, \hyg{1,-\eps,1-\eps; \frac{a}{s_1} \pm\iztilde}, \quad \text{where } a\in \mathbb{R}\!\setminus\! \lbrace 0\rbrace \,.
		\label{eq:first_and_third_limit_s1_to_zero}
	\end{equation}
To evaluate this limit, we first apply the inversion formula for the hypergeometric function derived in eq.\,\eqref{eq:2F1_inversion_formula} in the appendix, which yields
	\begin{align}
		\text{eq.\,\eqref{eq:first_and_third_limit_s1_to_zero}} \,=\, \lim_{s_1\rightarrow 0}\; \frac{1}{s_1} \left\lbrace - \hyg{1,\eps,1+\eps; \frac{s_1}{a}} \,+\, 1 \,+\, \left(\frac{-a}{s_1} \mp \iztilde\right)^{\!\eps}\frac{\pi\eps}{\sin(\pi\eps)} \right\rbrace .
	\end{align}
Imposing the condition $\eps <-1$, which is needed for convergence in case $s_1=0$, the second term vanishes. The remaining limit can be evaluated using L'H\^{o}pital's rule, 
	\begin{align}
		\text{eq.\,\eqref{eq:first_and_third_limit_s1_to_zero}} \,=\, \lim_{s_1\rightarrow 0}\; \left\lbrace -\, \frac{1}{a} \left.\frac{\dx}{\dx x}\,\hyg{1,\eps,1+\eps;x}\right|_{x=\frac{s_1}{a}}\right\rbrace \,=\, -\,\frac{1}{a}\,\frac{\eps}{1+\eps} \,.
	\end{align}
The derivative of the hypergeometric function was read off the series expansion given in eq{.}\,\eqref{eq:SeriesDefinition_2F1}. The second term in the general result also vanishes for $s_1\rightarrow 0$ since $\eps<-1$.
Combining everything, we find that taking the limit $s_1\rightarrow 0$ of the general result given in eq.\,\eqref{eq:D0_general_result_combined_factors} is equivalent to setting $s_1=0$ in eq.\,\eqref{eq:D0_xi_integrals}.

\subsection[$s_1=0$ and $s_2=0$]{\boldmath{$s_1=0$} and $\boldmath{s_2=0}$}
In this case, we set $s_2=0$ in eq.\,\eqref{eq:D0_s1_vanishing_integral} and obtain
\begin{equation}
	D_0\!\left(s_1=0,s_2=0,q^2=s_3\right) \,=\, -\frac{1}{\eps (1+\eps)} \frac{\Gamma(1+\eps) \Gamma^2(1-\eps)}{\Gamma(1-2\eps)}\, \frac{2}{s_3^2} \left(\frac{\mu^2}{-s_3-\iz}\right)^{\!\eps} .
\end{equation}
This is identical to the expression obtained by taking the limit $s_2\rightarrow 0$ of eq.\,\eqref{eq:D0_s1_vanishing}, i.e. taking the limits $s_1,s_2\rightarrow 0$ of eq.\,\eqref{eq:D0_general_result_combined_factors}.

\subsection[$s_3=0$]{\boldmath{$s_3=0$}}
In this case, eq.\,\eqref{eq:D0_xi_integrals} reads
	\begin{align}
		D_0\!\left(s_1=0,s_2,q^2=s_2\right) &=\, \mu^{2\eps}\, \frac{\Gamma(2+\eps) \Gamma^2(-\eps)}{\Gamma(-2\eps)} \int_0^1\dx\xi_1 \int_0^1\dx\xi_2\, \left[- s_1 \xi_1 - s_2\xi_2 -\iz\right]^{-\eps-2} \nonumber
		\\
		&=\,  \frac{1}{\eps}\frac{\Gamma(1+\eps) \Gamma^2(1-\eps)}{\Gamma(1-2\eps)}\, \frac{2}{s_2} \mu^{2\eps} \nonumber\\
		&\hphantom{=}\times \int_0^1\dx\xi_1 \left\lbrace \left[ -s_1\xi_1 -\iz\right]^{-\eps-1} - \left[ -s_1\xi_1 -s_2 -\iz\right]^{-\eps-1} \right\rbrace ,
	\end{align}
where the $\xi_2$-integration was performed elementarily. Note that $\eps < -1$ was imposed for convergence. Next, we substitute ${\xi_1\rightarrow \zeta = -s_1\xi_1}$ in the first term and ${\xi_1\rightarrow \zeta = -s_1\xi_1-s_2}$ in the second term, and subsequently split the latter integral into two separate integrals,
	\begin{align}
		D_0\!\left(s_1,s_2,q^2=s_1+s_2\right) &=\, \frac{1}{\eps}\frac{\Gamma(1+\eps) \Gamma^2(1-\eps)}{\Gamma(1-2\eps)}\, \frac{2}{s_1s_2}\, \mu^{2\eps} \left\lbrace -\int_0^{-s_1}\dx\zeta \left[ \zeta-\iz\right]^{-\eps-1} \right. \nonumber
		\\
		&\hphantom{=} \left. -\int_0^{-s_2}\dx\zeta \left[\zeta -\iz\right]^{-\eps-1} +\int_0^{-q^2}\dx\zeta\left[\zeta -\iz\right]^{-\eps-1} \right\rbrace .
	\end{align}
Note that the last integral vanishes for $q^2=0$. Evaluating the remaining $\zeta$-integrals yields
	\begin{align}
		D_0\!\left(s_1,s_2,q^2=s_1+s_2\right) =\, &\frac{1}{\eps^2}\frac{\Gamma(1+\eps) \Gamma^2(1-\eps)}{\Gamma(1-2\eps)}\, \frac{2}{s_1s_2} \nonumber
		\\
		& \times \left\lbrace \left(\frac{\mu^2}{-s_1-\iz}\right)^{\!\eps} + \left(\frac{\mu^2}{-s_2-\iz}\right)^{\!\eps} - \left(\frac{\mu^2}{-q^2-\iz}\right)^{\!\eps} \right\rbrace ,
	\end{align}
which can be analytically continued beyond $\eps<-1$ which was originally required for convergence. One can easily see that this result is equivalent to setting $s_3=0$ in the general result as given in eq.\,\eqref{eq:D0_general_result_combined_factors}, since the resulting hypergeometric functions evaluate to ${\hyg{1,-\eps,1-\eps;0} = 1}$ according to eq.\,\eqref{eq:SeriesDefinition_2F1}.

\subsection[$s_1=s_3=0$ or $s_2=s_3=0$]{\boldmath{$s_1=s_3=0$} or \boldmath{$s_2=s_3=0$}}
As above, we can without loss of generality consider $s_1=s_3=0$. In this case, eq.\,\eqref{eq:D0_xi_integrals} yields
	\begin{align}
		D_0\!\left(s_1,s_2,q^2=s_1+s_2\right) &=\, \mu^{2\eps}\, \frac{\Gamma(2+\eps) \Gamma^2(-\eps)}{\Gamma(-2\eps)} \int_0^1\dx\xi_1 \int_0^1\dx\xi_2\, \left[(- s_2-\iz)\xi_2\right]^{-\eps-2}
		\nonumber \\
		&=\, \frac{1}{\eps} \frac{\Gamma(1+\eps) \Gamma^2(1-\eps)}{\Gamma(1-2\eps)}\, \frac{2}{s_2^2} \left(\frac{\mu^2}{-s_2-\iz}\right)^{\!\eps}\,.
	\end{align}
As in the previous cases, one has to impose $\eps<-1$ for convergence before analytically continuing to larger $\eps$. 

Let us compare this expression to the corresponding limit of the general result as given in eq.\,\eqref{eq:D0_general_result_combined_factors}. For $s_3=0$, the hypergeometric functions evaluate to ${\hyg{1,-\eps,1-\eps;0} = 1}$. The limit remaining to be taken is
	\begin{align}
		\lim_{s_{1},s_{3}\rightarrow 0} D_0\!\left(s_1,s_2,q^2\right) \,&=\, \frac{1}{\eps^2} \frac{\Gamma(1+\eps)\, \Gamma^2(1-\eps)}{\Gamma(1-2\eps)}\, \frac{2}{s_2}\nonumber \\
		&\times\,
		 \left\lbrace \lim_{s_1\rightarrow 0} \frac{\left[\frac{\mu^2}{-s_2-\iz}\right]^{\eps} \,- \left[\frac{\mu^2}{-s_1-s_2-\iz}\right]^{\eps}}{s_1} \right.		
		\;\left. +\, \lim_{s_1\rightarrow 0} \frac{\left[\frac{\mu^2}{-s_1-\iz}\right]^{\eps}}{s_1} \right\rbrace .
		 \label{D0_lim_s1s3}
	\end{align}
Imposing the condition $\eps <-1$, which is needed for convergence in case $s_1=0$, the second term vanishes. Applying L'H\^{o}pital's rule to the first limit, we find that the limit ${s_{1},s_{3}\rightarrow 0}$ of the general result given in eq.\,\eqref{eq:D0_general_result_combined_factors} is equivalent to setting $s_1=s_3=0$ in eq.\,\eqref{eq:D0_xi_integrals}.

\subsection[$s_1=s_2=s_3=0$]{\boldmath{$s_1=s_2=s_3=0$}}\label{sec:All_Mandelstam_Vanish}
Setting all Mandelstam variables to $0$  in eq.\,\eqref{eq:D0_xi_integrals} and requiring $\eps<-2$ for convergence yields
	\begin{equation}
		D_0\!\left(s_1=0,s_2=0,q^2=0\right) \,=\, \mu^{2\eps}\, \frac{\Gamma(2+\eps) \Gamma^2(-\eps)}{\Gamma(-2\eps)} \int_0^1\dx\xi_1 \int_0^1\dx\xi_2\, \left[0\right]^{-\eps-2} \,=\, 0 \,.
	\end{equation}
This integral also has to vanish since the integrand has a mass dimension of $-\eps-2$, but does not depend on any mass scale. Taking the limit ${s_{1,2,3}\rightarrow 0}$ of the general result given in eq.\,\eqref{eq:D0_general_result_combined_factors} leads to the same conclusion.

\section{Gauss hypergeometric function \boldmath{$\hyg{1,-\eps,1-\eps;z}$}}\label{app:2F1}
For $|z|<1$ the Gauss hypergeometric function is defined by the series \citep[eq.\,(15.1.1)]{Abramovitz_Stegun}
\begin{align}
\hyg{a,b,c;z}=\sum_{n=0}^\infty\frac{(a)_n(b)_n}{(c)_n}\,\frac{z^n}{n!}\,,
\label{eq:SeriesDefinition_2F1}
\end{align}
where $(a)_n=\frac{\Gamma(a+n)}{\Gamma(a)}$ is the Pochhammer symbol.
It is analytically continued by the integral representation \citep[eq.\,(15.3.1)]{Abramovitz_Stegun}
	\begin{align}
		\hyg{a,b,c;z} \,\equiv\, &\frac{\Gamma(c)}{\Gamma(b)\, \Gamma(c-b)} \int_0^1\dx t\, t^{b-1}\,(1-t)^{c-b-1}\, (1-zt)^{-a} \,,  \label{eq:IntegralRepresentation_2F1}
		\\
		&\text{where}\; \mathrm{Re}(c)>\mathrm{Re}(b)>0\,,\; z \in \mathbb{C}\!\setminus\! \mathbb{R}_{\geq 1}\,. \nonumber
	\end{align}
Note that the integral may be ill-defined for $z>1$, since in this case the term ${(1-zt)}$ changes sign within the range of integration. This change of sign can, in general, give rise to a discontinuity or branch cut. In the following, we discuss selected properties (including the branch cut structure) of the specific hypergeometric function $\hyg{1,-\eps,1-\eps;z}$ which appears in the internal-massless single off-shell scalar box integral.

\subsection[Epsilon expansion of $\hyg{1,-\eps,1-\eps;z}$]{Epsilon expansion of \boldmath{$\hyg{1,-\eps,1-\eps;z}$}}The $\eps$-expansion of this hypergeometric function was derived in ref.\,\citep[eq.\,(B.28)]{FabiValery},
	\begin{align}
		\hyg{1,-\eps,1-\eps;z} \,=\, 1 \,-\, \sum_{n=1}^{\infty} \eps^n \,\Li{n}{z}.
		\label{eq:2F1_eps_expansion}
	\end{align}
Since the polylogarithms $\Li{n}{z}$ have a branch cut along the positive real axis greater than 1 \citep{Lewin_Polylogs_Associated_Functions}, the hypergeometric function also has a branch cut there. The branch cut structure will be discussed in section \ref{sec:2F1_branch_cut_structure}.

\subsection[Inversion relation for $\hyg{1,-\eps,1-\eps;z}$]{Inversion relation for \boldmath{$\hyg{1,-\eps,1-\eps;z}$}}
An equation relating the hypergeometric function $\hyg{1,-\eps,1-\eps;z}$ to another hypergeometric function with argument $\frac{1}{z}$ is useful because it allows us to transform the argument away from the branch cut. To find this inversion relation, we write the hypergeometric function in terms of its $\eps$-expansion and invert the appearing polylogarithms using the inversion relation for the polylogarithm given in ref.\,\citep[eq.\,(7.20)]{Lewin_Polylogs_Associated_Functions},
	\begin{align}
		\Li{n}{z} \,+\, (-1)^n\,\Li{n}{\frac{1}{z}} \,=\, -\frac{\ln^n(-z) }{n!} \,+\, 2\sum_{k=1}^{\lfloor n/2\rfloor} \frac{\ln^{n-2k}(-z)}{(n-2k)!}\,\Li{2k}{-1}\,,
	\end{align}
which holds for $z\in\mathbb{C}\setminus[0,\infty)$.\footnote{For reference, the corresponding equation on the branch cut, i.e. for $x\in (0,\infty)$, is
\begin{align*}
\Li{n}{x\pm\iz}\,+\,(-1)^{n}\,\Li{n}{\frac{1}{x\pm\iz}}\,=\,-\frac{\ln^n(x)}{n!}\,+\,2\sum_{k=1}^{\lfloor n/2\rfloor} \zeta_{2k} \frac{\ln^{n-2k}(x)}{(n-2k)!}\,\pm\,\iu \pi\,\frac{\ln^{n-1}(x)}{(n-1)!}\,.
\end{align*}
} Plugging the second term on the left into eq.\,\eqref{eq:2F1_eps_expansion} leads to the closely related {$\eps$-expansion} of $\hyg{1,\eps,1+\eps;\frac{1}{z}}$, while plugging the second term on the right into eq.\,\eqref{eq:2F1_eps_expansion} leads to the series expansion of an exponential function, such that
	\begin{align}
		&\hyg{1,-\eps,1-\eps;z} +\, \hyg{1,\eps,1+\eps;\frac{1}{z}}\nonumber \\
		=&\,
		1 \,+\, (-z)^{\eps}
		-\, 2 \sum_{n=1}^{\infty}\eps^n \sum_{k=1}^{\lfloor n/2\rfloor}\frac{\ln^{n-2k}(-z)}{(n-2k)!}\Li{2k}{-1} \,.
	\end{align}
To simplify the term on the second line of this expression, we interchange the two sums and subsequently shift the summation index to $N=n-2k$, such that the $N$-sum becomes an exponential function,
	\begin{equation}
		\sum_{n=1}^{\infty}\eps^n \sum_{k=1}^{\lfloor n/2\rfloor}\frac{\ln^{n-2k}(-z)}{(n-2k)!}\Li{2k}{-1} \,=\, \sum_{k=1}^{\infty}\Li{2k}{-1}\eps^{2k} \underbrace{\sum_{N=0}^{\infty}\frac{\left(\eps\ln(-z)\right)^{N}}{N!}}_{=(-z)^{\eps}} \,.
	\end{equation}
Thus, we obtain
	\begin{align}
		\hyg{1,-\eps,1-\eps;z} +\, \hyg{1,\eps,1+\eps;\frac{1}{z}} \,&=\,
		1 \,+\, (-z)^{\eps} \left[1 \,-\, 2\,\sum_{k=1}^{\infty}\Li{2k}{-1}\eps^{2k} \right].
	\end{align}
The remaining polylogarithm constants can be written as \citep[eq.\,(7.29)]{Lewin_Polylogs_Associated_Functions}
	\begin{align}
		\Li{2k}{-1} \,=\, - (-1)^{k-1}\left( 2^{2k-1}-1\right) \BernoulliB{2k}\frac{\pi^{2k}}{(2k)!} \,,
		\label{eq:Polylog_Bernoulli_numbers}
	\end{align}
where $\BernoulliB{2k}$ are the Bernoulli numbers, which are given by the coefficients in the following series expansion\footnote{Note that this modern definition of the Bernoulli numbers differs slightly from the older definition given in ref.\,\citep[eq.\,(7.23)]{Lewin_Polylogs_Associated_Functions}, the relation cited in eq.\,\eqref{eq:Polylog_Bernoulli_numbers} was adjusted accordingly.}
	\begin{align}
		\frac{t}{e^t-1} \,=\, \sum_{n=0}^{\infty} \BernoulliB{n} \frac{t^n}{n!} \,.
	\end{align}
Using this, we find that the term in square brackets is the series representation of $\pi \eps \csc(\pi\eps) = \pi\eps / \sin(\pi\eps)$ \citep[eq.\,(4.3.68)]{Abramovitz_Stegun},
	\begin{align}
		1 \,-\, 2\,\sum_{r=1}^{\infty}\Li{2r}{-1}\eps^{2r}  \,&=\, 1 \,+\, \sum_{r=1}^{\infty} \frac{(-1)^{r-1}\, 2 \left(2^{2r-1} -1\right) \BernoulliB{2r}}{(2r)!} (\pi\eps)^{2r} \,=\, \frac{\pi\eps}{\sin(\pi\eps)} \,.
	\end{align}
Hence the inversion formula for the hypergeometric function is
	\begin{align}
		\hyg{1,-\eps,1-\eps;z} +\, \hyg{1,\eps,1+\eps;\frac{1}{z}} =\, 1 \,+\, (-z)^{\eps}\, \frac{\pi\eps}{\sin(\pi\eps)} \,.
		\label{eq:2F1_inversion_formula}
	\end{align}

\subsection[Branch cut structure of $\hyg{1,-\eps,1-\eps;z}$]{Branch cut structure of \boldmath{$\hyg{1,-\eps,1-\eps;z}$}}\label{sec:2F1_branch_cut_structure}
The hypergeometric function $\hyg{1,-\eps,1-\eps;z}$ has a branch cut along the positive real axis starting at $1$. This is explicitly seen through the inversion formula derived in eq.\,\eqref{eq:2F1_inversion_formula}. If we take $x\in \mathbb{R}_{>1}$ and $\iz$ as an infinitesimal imaginary part, then we have 
	\begin{align}
		\hyg{1,-\eps,1-\eps; x\pm \iz} \,=\, -\, \hyg{1,\eps,1+\eps;\frac{1}{x}} \,+\, 1 \,+\, (-x\mp\iz)^{\eps} \frac{\pi\eps}{\sin(\pi\eps)} \,. 
	\end{align}
Note that the imaginary part could be dropped in the hypergeometric function on the right-hand side since the function is well-defined for arguments $1/x < 1$. Explicitly writing the factor $(-x\mp\iz)^{\eps}$ in terms of its real and imaginary part, we obtain
	\begin{align}
		\hyg{1,-\eps,1-\eps; x\pm \iz} \,=\, -\, \hyg{1,\eps,1+\eps;\frac{1}{x}} \,+\, 1 \,+\, \pi\eps \cot(\pi\eps)\, x^{\eps} \,\mp \iu \pi\eps\, x^{\eps}\,. 
	\end{align}

\section{Single-valued polylogarithms}
\label{app:SingleValuedPolylogs}
\subsection{Single-valued polylogarithms in the literature}
There are several definitions of so called \textit{single-valued polylogarithms} in the literature. Their common feature is that they do not have branch cuts opposed to regular polylogarithms. Depending on the intended use, different versions prove themselves to be most suitable \citep{Zhao2016}.

Bloch and Wigner devised an important version of a single-valued dilogarithm \citep{Bloch2011}
\begin{align}
D_2(z)\,=\,\mathrm{Im}\left[\Li{2}{x}\right]+\mathrm{arg}(1-z)\ln|z|\,,
\end{align}
defined on $\mathbb{C}\setminus\lbrace 0,1\rbrace$. This was generalized to polylogarithms by Ramakrishnan \citep{Ramakrishnan1986}. With a slight modification for odd $n$ due to Wojtkowiak \citep{Wojtkowiak1989} and the notation $\mathfrak{R}_n=\mathrm{Re}$\,/\,$\mathrm{Im}$ if $n$ is odd\,/\,even, his definition reads
	\begin{align}
		\LiSVRamakrishnan{n}{z}=\mathfrak{R}_n\left[\sum_{k=0}^{n-1}\frac{(-1)^k \ln^k|z|}{k!}\Li{n-k}{z}-\frac{(-1)^n\ln^n|z|}{2 n!}\right].
	\end{align}
Similarly Zagier proposed \citep[eq.\,(33)]{Zagier1991}
	\begin{align}
		\LiSVZagier{n}{z}=\mathfrak{R}_n\left[\sum_{k=0}^{n-1}\frac{2^k \BernoulliB{k}}{k!}\ln^k|z|\,\Li{n-k}{z}-\frac{2^{n-1}\BernoulliB{n}}{n!}\ln^n|z|\right].
	\end{align}
Both are single-valued real-analytic functions from $\mathbb{C}\setminus\lbrace 0,1\rbrace$ to $\mathbb{R}$.
Zagier's version satisfies the functional equations of the polylogarithm in a clean fashion without product terms. However, both Ramakrishnan's and Zagier's functions are identically zero on $\mathbb{R}$ for even $n$. Hence, they are not suitable for our purpose.

Also in an attempt to get rid of product terms in the functional equations, Lewin defined his version of polylogarithms by \citep[eq.\,(3.19)]{Lewin1991}
	\begin{align}
		\LiSVLewin{n}{x}=\sum_{k=0}^{n-1}\frac{(-1)^k}{k!}\,\ln^k|x|\,\Li{n-k}{x}+\frac{(-1)^{n-1}}{n!}\ln^{n-1}|x|\ln(1-x)
		\label{eq:Definiton Lewin SVP x<1}
	\end{align}
for $x<1$, as a generalization of Rogers' dilogarithm $\LiSVLewin{2}{x}$ \citep[eq.\,(1)]{Rogers1907}.
They are related to Kummer's functions \citep[p.\,330]{Kummer1840}
${
\Lambda_n(x)=\int_0^x\dx t\, \frac{\ln^{n-1}|t|}{1+t}
}$
by \citep[eq.\,(3.18)]{Lewin1991}
\begin{align}
\LiSVLewin{n}{x}=\frac{(-1)^{n}}{(n-1)!}\,\Lambda_n(-x)+\frac{(-1)^{n-1}}{n!}\ln^{n-1}|x|\ln(1-x)\,.
\label{eq: Relation Lewin Kummer}
\end{align}
Kummer's functions are real-valued for real arguments, which makes them single-valued polylogarithms predating Bloch and Wigner's construction by well over a century.
We observe from eq.\,\eqref{eq: Relation Lewin Kummer} that $\LiSVLewin{n}{x}$ can be continued to a single-valued function for $x>1$ by replacing $\ln(1-x)\rightarrow\ln|1-x|$ in definition \eqref{eq:Definiton Lewin SVP x<1}.
\subsection{Definition and basic properties of continuous single-valued polylogarithms}
\begin{figure}[htp!]
		\centering
		\includegraphics[width=0.45\textwidth]{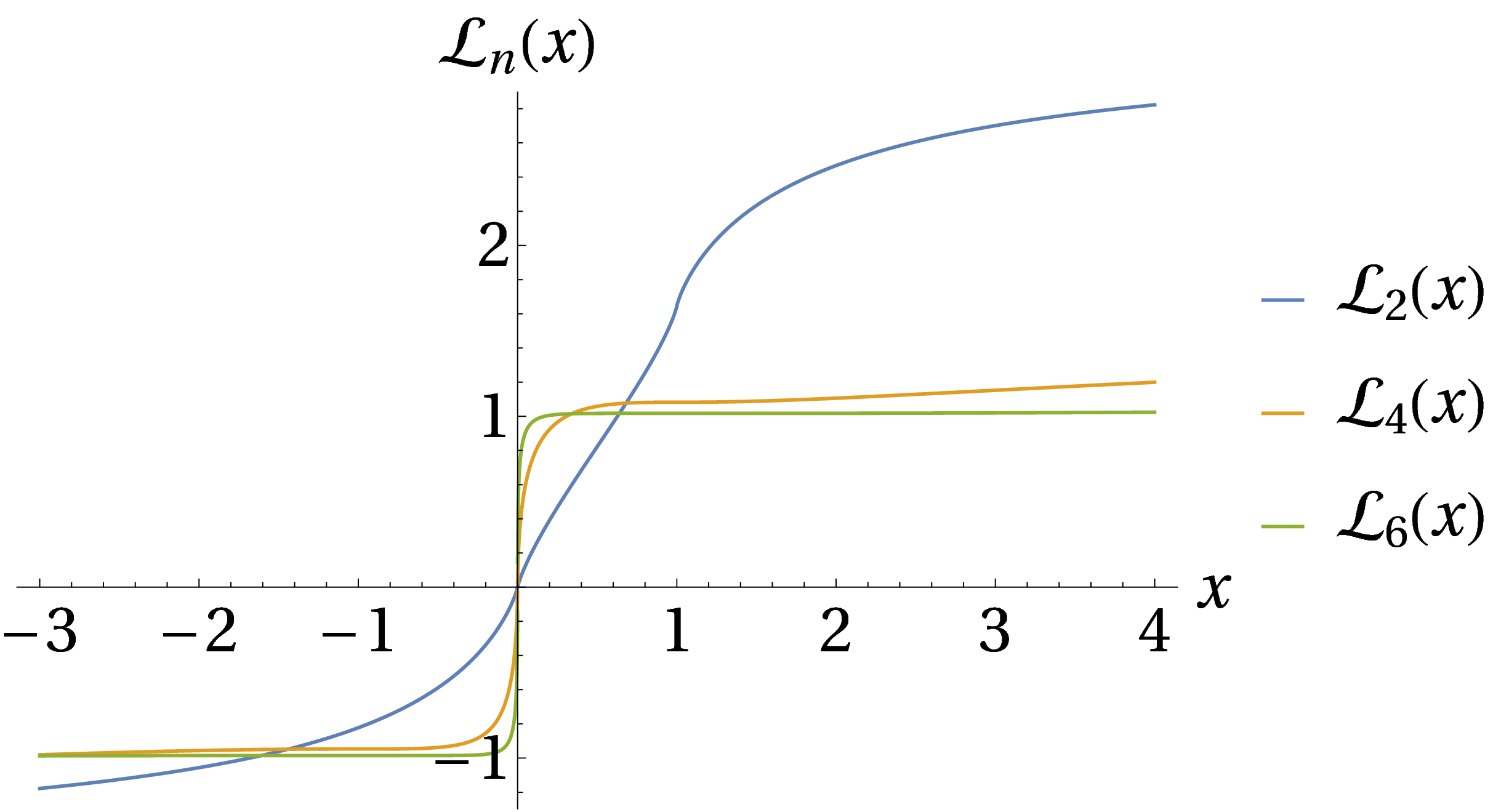}
		\includegraphics[width=0.45\textwidth]{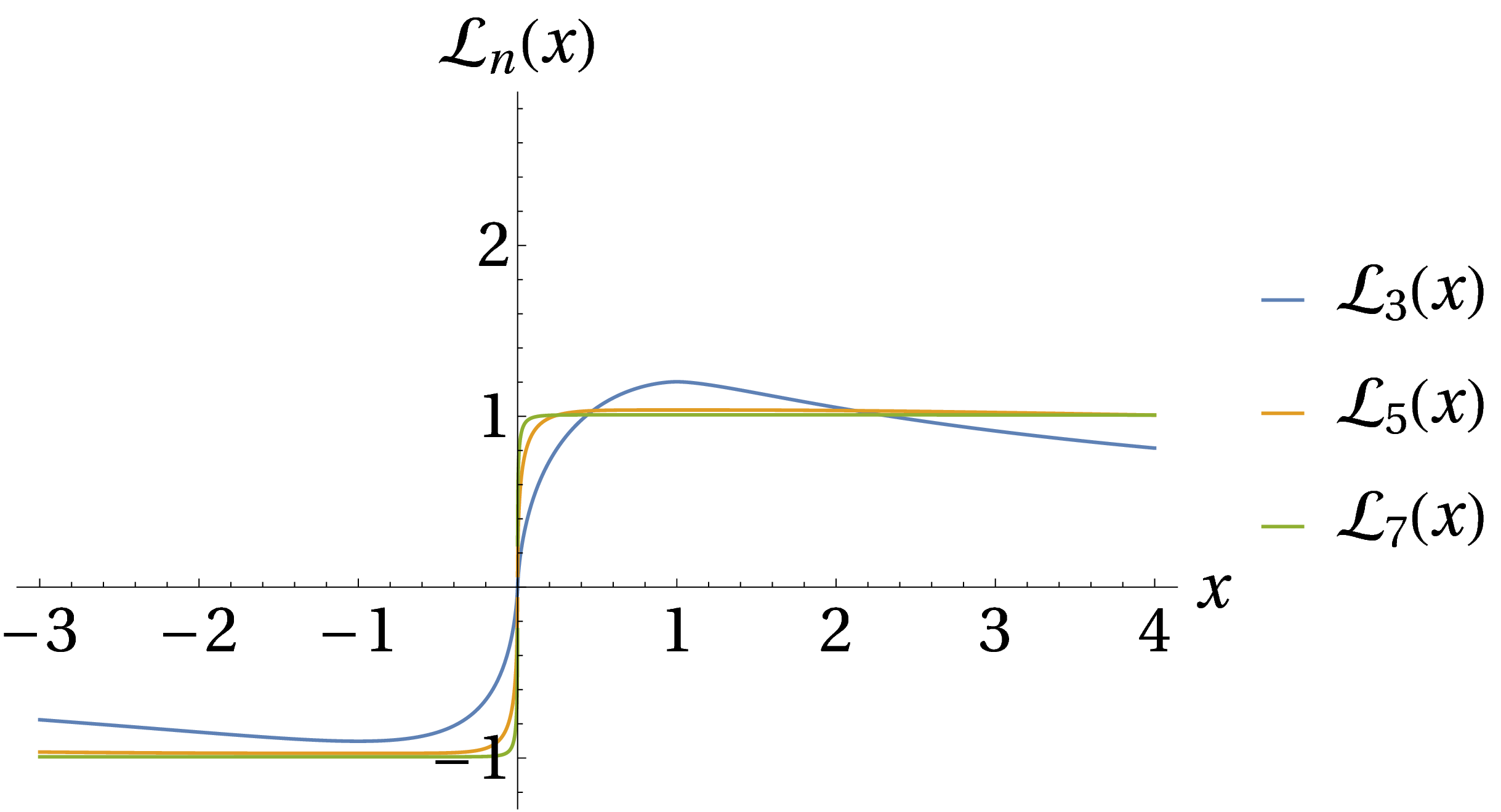}
		\caption{Plots of the single-valued polylogarithms defined in eq.\,\eqref{eq:Definiton SVP}, on the left for even $n$, on the right for odd $n$.}
		\label{fig:Plot SVP}
\end{figure}

We define for $n\in\mathbb{N}_{\geq 2}$ and real $x$ a version of single-valued polylogarithms by
	\begin{tcolorbox}[colback=green!10!white]
		\begin{align}
			\LiSV{n}{x} \,\equiv\, \sum_{k=0}^{n-1}\frac{(-1)^k}{k!}\,\ln^k|x|\,\Li{n-k}{x}+\frac{(-1)^{n-1}}{n!}\ln^{n-1}|x|\ln|1-x|\,.
			\label{eq:Definiton SVP}
		\end{align}
	\end{tcolorbox}
\noindent{}For $x>1$ the polylogarithms are taken on their principal branch, which is inherited from the principal branch of the logarithm, i.e. ${\Li{n}{x}\equiv\Li{n}{x-\iz}}$. As a result we have a function of a real variable $x$ with the following properties:
	\begin{itemize}
		\item[(a)] $\LiSV{n}{x}$ is single-valued, i.e. there is no branch cut and no imaginary part, in contrast to $\Li{n}{x}$.
		\item[(b)] $\LiSV{n}{x}$ is continuous for all $x\in\mathbb{R}$.
		\item[(c)] $\LiSV{n}{x}$ is bounded on $\mathbb{R}$, in contrast to $\Li{n}{x}$.
		\item[(d)] $\LiSV{n}{x}$ satisfies \textit{clean} versions of the functional equations of $\Li{n}{x}$, i.e. without product terms, as for Lewin's and Zagier's functions.
	\end{itemize} 
These properties make the function $\LiSV{n}{x}$ well-suited for various applications, where one is interested in having a well-behaved version of polylogarithms with real arguments. Plots of the first few single-valued polylogarithms are shown in figure \ref{fig:Plot SVP}.

The qualitative behaviour of the functions is different for even and odd $n$. Here, $\zeta_n=\zeta(n)=\sum_{k=1}^\infty k^{-n}$ are the integer values of the Riemann zeta function.
	\begin{itemize}
		\item[]\textbf{For even \boldmath{$n$}:} $\LiSV{n}{x}$ is monotonously increasing with $\LiSV{n}{-\infty}=2(2^{1-n}-1)\zeta_n$ and $\LiSV{n}{\infty}=2\zeta_n$. Special values are $\LiSV{n}{-1}=(2^{1-n}-1)\zeta_n$, $\LiSV{n}{0}=0$, $\LiSV{n}{1}=\zeta_n$. These functions are discontinuous at infinity, meaning $\LiSV{n}{\infty}\neq\LiSV{n}{-\infty}$. Therefore functional equations mapping infinity to $0$ or $1$ will be discontinuous at those points.
		\item[]\textbf{For odd \boldmath{$n$}:} It is $\LiSV{n}{-\infty}=0$, $\LiSV{n}{x}$ is decreasing in $(-\infty,-1)$ to a minimum value of $\LiSV{n}{-1}=(2^{1-n}-1)\zeta_n$, increases in $(-1,1)$, with $\LiSV{n}{0}=0$, to a maximum of $\LiSV{n}{1}=\zeta_n$ and monotonously falls in $(1,\infty)$ to $\LiSV{n}{\infty}=0$. Since ${\LiSV{n}{\infty}=\LiSV{n}{-\infty}}$ these functions are continuous at infinity. Therefore functional equations mapping infinity to $0$ or $1$ are continuous.
	\end{itemize}
	
\noindent{}For both odd and even $n$, $\LiSV{n}{x}$ is real analytic on $\mathbb{R}\backslash\lbrace 0,1 \rbrace$. At $x=0$ its derivative diverges logarithmically, at $x=1$ it is $(n-2)$-times differentiable, the $(n-1)$th derivative diverges logarithmically at this point.

\subsection{Functional equations of continuous single-valued polylogarithms}
\label{sec:Functional equations of continuous single-valued polylogarithms}
The functional equations satisfied by single-valued polylogarithms are inherited from the corresponding polylogarithms \citep{Lewin_Polylogs_Associated_Functions}. Single-valued polylogarithms obey these equations in a \textit{cleaner} fashion, where all product terms vanish.
There are equations satisfied for all $n$ and additional identities for small values of $n$. We explicitly give the identities for di- and trilogarithm, identities for higher order polylogarithms can be found in \citep{Lewin_Polylogs_Associated_Functions,Lewin1991}.

\subsubsection{Equations of all single-valued polylogarithms}
For all $n$ there is an inversion relation and a duplication formula.
	\begin{itemize}
		\item[(i)] \textbf{Inversion relation}:
		\begin{align}
			\LiSV{n}{x}+(-1)^n\LiSV{n}{\frac{1}{x}}=\left\lbrace
			\begin{array}{cl} 0 & \text{ for odd }n\,,\\
				2\zeta_n & \text{ for even }n\text{ and }x>0\,,\\ 2(2^{1-n}-1)\zeta_n & \text{ for even }n\text{ and }x<0\,.
			\end{array}\right.
			\label{eq:PolyLog inversion relation}
		\end{align}
		Alternatively we can write this in the suggestive form
		\begin{align}
			\LiSV{n}{x}+(-1)^n\LiSV{n}{\frac{1}{x}}=\LiSV{n}{\sgn{x}\infty}.
		\end{align}
		\item[(ii)] \textbf{Duplication formula}:
		\begin{align}
			\LiSV{n}{x}+\LiSV{n}{-x}-2^{1-n}\LiSV{n}{x^2}=0\,.
		\end{align}
	\end{itemize}

\subsubsection{Equations of the single-valued dilogarithm}
For the single-valued dilogarithm, defined as
	\begin{align}
		\LiSV{2}{x}=\Li{2}{x}+\ln|x|\ln(1-x)-\frac{1}{2}\ln|x|\ln|1-x|\,,
	\end{align}
there is the richest structure. We have identities connecting the arguments $x$, $1-x$, $\frac{x}{x-1}$, $\frac{1}{x}$, $\frac{1}{1-x}$ and $1-\frac{1}{x}$. Additionally there is the famous 2-variable 5-term relation discovered independently by Spence \citep[p.9]{Spence1809} and Abel \citep[XIV.(9)]{Abel1881}.
All discontinuities in these identities are induced by the mapping of the discontinuity at infinity.
\begin{itemize}
	\item[(i)] \textbf{Euler identity}:
	\begin{align}
		\LiSV{2}{x}+\LiSV{2}{1-x}=\zeta_2\,.
		\label{eq:Euler Li2}
	\end{align}
	\item[(ii)] \textbf{Landen identity}:
	\begin{align}
		\LiSV{2}{x}+\LiSV{2}{\frac{x}{x-1}}=\left\lbrace\begin{array}{cl}
			0 & \text{for }x<1\,,\\
			3\zeta_2 & \text{for }x>1\,.
		\end{array}\right.
		\label{eq:Landen Li2}
	\end{align}
	\item[(iii)] \textbf{Inversion identity}:
	\begin{align}
		\LiSV{2}{x}+\LiSV{2}{\frac{1}{x}}=\left\lbrace\begin{array}{cl}
			-\zeta_2 & \text{for }x<0\,,\\
			2\zeta_2 & \text{for }x>0\,.
		\end{array}\right.
		\label{eq:Inversion Li2}
	\end{align}
	\item[(iv)] \textbf{Inverted Euler identity:}
	\begin{align}
		\LiSV{2}{x}-\LiSV{2}{\frac{1}{1-x}}=\left\lbrace\begin{array}{cl}
			-\zeta_2 & \text{for }x<1\,,\\
			2\zeta_2 & \text{for }x>1\,.
		\end{array}\right.
		\label{eq:Inverted Euler Li2}
	\end{align}
	\item[(v)] \textbf{Inverted Landen identity:}
	\begin{align}
		\LiSV{2}{x}-\LiSV{2}{1-\frac{1}{x}}=\left\lbrace\begin{array}{cl}
			-2\zeta_2 & \text{for }x<0\,,\\
			\zeta_2 & \text{for }x>0\,.
		\end{array}\right.
		\label{eq:Inverted Landen Li2}
	\end{align}
	\item[(vi)] \textbf{Abel identity:}
	\begin{align}
		&\LiSV{2}{x}+\LiSV{2}{y}+\LiSV{2}{\frac{1-x}{1-x y}}+\LiSV{2}{1-x y}+\LiSV{2}{\frac{1-y}{1-x y}}=\pm 3\zeta_2\,,
		\label{eq:Abel Li2}
	\end{align}
	where the sign on the right is negative if $1-xy<0$ and $x,y<0$, else positive.
\end{itemize}

\subsubsection{Equations of the single-valued trilogarithm}
\label{sec:Equations of the single-valued trilogarithm}
Besides the inversion relation, the single-valued trilogarithm, defined as
	\begin{align}
		\LiSV{3}{x}=\Li{3}{x}-\ln|x|\,\Li{2}{x}-\frac{1}{2}\ln^2|x|\ln(1-x)+\frac{1}{6}\ln^2|x|\ln|1-x|\,,
	\end{align}
satisfies a single-variable identity connecting three terms and a three-variable 22 term relation discovered by Goncharov \citep{Goncharov1995}. From the latter there follow a plethora of two variable identities as special cases.
Since the single-valued trilogarithm is continuous at infinity, these identities are continuous.
\begin{itemize}
	\item[(i)] \textbf{Inversion relation:}
	\begin{align}
		\LiSV{3}{x}-\LiSV{3}{\frac{1}{x}}=0\,.
	\end{align}
	\item[(ii)] \textbf{Landen identity:}
	\begin{align}
		\LiSV{3}{x}+\LiSV{3}{1-x}+\LiSV{3}{\frac{x}{x-1}}=\zeta_3\,.
	\end{align}
	\item[(iii)] \textbf{Goncharov identity:}
	\begin{align}
		\LiSV{3}{x y z}+&\bigoplus_{\mathrm{Cyc}(x,y,z)}
		\left[\LiSV{3}{x}-\LiSV{3}{xy}+\LiSV{3}{\frac{x(1-y)}{x-1}}+\LiSV{3}{\frac{y(1-x)}{y-1}}\right.
		\nonumber\\
		+&\left.\LiSV{3}{\frac{1-x}{1-xyz}}+\LiSV{3}{\frac{x y(1-z)}{1-xyz}}-\LiSV{3}{\frac{x(1-y)(1-z)}{(x-1)(1-xyz)}}\right]=3\zeta_3\,,
	\end{align}
	where cyclic summation for a function $f(x,y,z)$ is defined as
	\begin{align}
		\bigoplus_{\mathrm{Cyc}(x,y,z)}f(x,y,z)=f(x,y,z)+f(y,z,x)+f(z,x,y)\,.
	\end{align}
\end{itemize}

\section{Example: Regularizing kinematic limits for Drell-Yan differential in rapidity}\label{app:Regularizing_kinematic_limits}
\begin{figure}[htp!]
		\centering
		\includegraphics[width=0.8\textwidth]{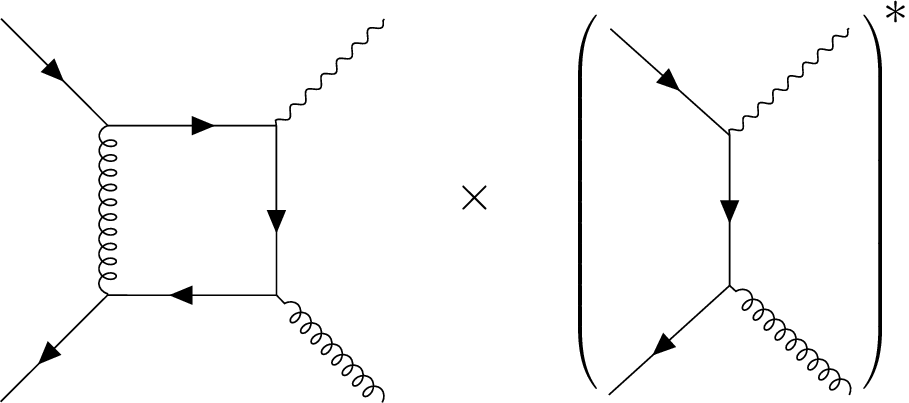}
		\caption{NNLO contribution to $q\bar{q}\rightarrow \gamma^\ast$ from interference of box and tree-level amplitudes. There are three different boxes and two different tree-level diagrams for this process.}
		\label{fig:DY NNLO diagrams}
\end{figure}

In this appendix, we illustrate the regularization of kinematic divergences occurring in the box integral using the Drell-Yan process differential in momentum transfer $Q^2$ and rapidity $y$ at $\mathcal{O}(\alpha\alpha_s^2)$ as an example.
This process was first calculated to NNLO in \citep{Anastasiou2003} using the reversed unitarity method. 
Instead, for the purpose of our example, we consider the NNLO calculation using the traditional treatment of the phase space integral as was done for the NLO calculation in \citep{Altarelli1979}.
We will explicitly see that orders beyond $\eps^0$ of $D_0$ are required and the full $\eps$-dependence of any terms divergent in the kinematic limit is needed here.
To illustrate these points it is sufficient to look at the contribution to $\frac{\dx \sigma}{\dx Q^2\dx y}$ from the quark-antiquark process. At NNLO the box integral appears through the interference term of one-loop and tree level diagrams for $q\bar{q}\rightarrow \gamma^\ast g$, which are depicted in figure \ref{fig:DY NNLO diagrams}.

The two particle phase space in $d=4-2\eps$ dimensions is given by (see eqs.\,(79) and (80) in \citep{Altarelli1979})
\begin{align}
\dx\mathrm{PS}_2=\frac{1}{8\pi}\left(\frac{4\pi}{Q^2}\right)^\eps \frac{z^\eps (1-z)^{1-2\eps}}{\Gamma(1-\eps)}\int_0^1\dx y\,y^{-\eps}(1-y)^{-\eps}\,,
\end{align}
with
\begin{align}
s=\frac{Q^2}{z}\,,\quad\,t=\frac{-Q^2}{z}(1-z)(1-y)\,,\quad\,u=-\frac{Q^2}{z}(1-z)y\,.
\end{align}
For simplicity, we only perform the regularization for $y\rightarrow 1$ in the following. Treatment of $y\rightarrow 0$ and $z\rightarrow 1$ would be analogous.
The relevant part of the phase space in the limit $y\rightarrow 1$ is $(1-y)^{-\eps}$. Calculating the interference terms of the type depicted in figure \ref{fig:DY NNLO diagrams}, we find that the scalar box integral contributes through

\begin{itemize}
\item[(i)]
$(1-y)^{-\eps} D_0\!\left(\frac{Q^2}{z},-\frac{Q^2}{z}(1-z)(1-y),Q^2\right)\,f_1(y,\eps)$\,,
\item[(ii)]
$(1-y)^{-\eps} D_0\!\left(\frac{Q^2}{z},-\frac{Q^2}{z}(1-z)y,Q^2\right)\,f_2(y,\eps)$\,,
\item[(iii)]
$(1-y)^{-1-\eps} D_0\!\left(\frac{Q^2}{z},-\frac{Q^2}{z}(1-z)y,Q^2\right)\,f_3(y,\eps)$\,,
\item[(iv)]
$(1-y)^{-\eps} D_0\!\left(-\frac{Q^2}{z}(1-z)y,-\frac{Q^2}{z}(1-z)(1-y),Q^2\right)\,f_4(y,\eps)$\,,
\end{itemize}
where the $f_i$ are functions finite in the limit $y\rightarrow 1$ order by order in $\eps$. The behavior of the scalar box functions $D_0$ for $y\rightarrow 1$ can be obtained from eq.\,\eqref{eq:D0 result with theta functions}. We find
\begin{align}
D_0(s,t,Q^2)&\sim \frac{1}{s t}\left|\frac{u \mu^2}{s t}\right|^\eps \left[\mathfrak{F}\!\left(\eps;-\frac{u}{s}\right)+\e^{\iu \pi\eps}\,\mathfrak{F}\!\left(\eps;-\frac{u}{t}\right)-\e^{\iu \pi\eps}\,\mathfrak{F}\!\left(\eps;-\frac{u Q^2}{st}\right)\right]
\nonumber\\
&\sim (1-y)^{-1-\eps}\left[\mathfrak{F}\!\left(\eps;(1-z)y\right)+\e^{\iu \pi\eps}\,\mathfrak{F}\!\left(\eps;-\frac{y}{1-y}\right)-\e^{\iu \pi\eps}\,\mathfrak{F}\!\left(\eps;-\frac{yz}{1-y}\right)\right],
\end{align}
where $\sim$ means that we dropped factors non-singular for $y\rightarrow 1$.
In the limit $y\rightarrow 1$ the argument of the first $\mathfrak{F}$ stays finite, for the latter two the argument tends to $-\infty$. According to eq.\,\eqref{eq:Ffrak limiting behavior -infty}, $\mathfrak{F}$ is finite order by order in $\eps$ in this case. Hence, we can write
\begin{align}
D_0(s,t,Q^2)=(1-y)^{-1-\eps} A_{st}(y)\,,
\label{eq:D0st behavior}
\end{align}
where $A_{st}(y)$ collects all factors which stay finite in the limit $y\rightarrow 1$.
Similarly,
\begin{align}
D_0(t,u,Q^2)&\sim\frac{1}{t u}\left|\frac{s \mu^2}{t u}\right|^\eps \left[\mathfrak{F}\!\left(\eps;-\frac{s}{t}\right)+\mathfrak{F}\!\left(\eps;-\frac{s}{u}\right)-\e^{\iu \pi\eps}\,\mathfrak{F}\!\left(\eps;-\frac{s Q^2}{t u}\right)\right]
\nonumber\\
&\sim (1-y)^{-1-\eps}\left[\mathfrak{F}\!\left(\eps;\frac{1}{(1-z)(1-y)}\right)+\,\mathfrak{F}\!\left(\eps;-\frac{1}{(1-z)y}\right)\right.\nonumber\\
&\hphantom{\sim (1-y)^{-1-\eps}}\,\,\left.-\e^{\iu \pi\eps}\,\mathfrak{F}\!\left(\eps;-\frac{1}{(1-z)^2 y (1-y)}\right)\right],
\end{align}
where again the bracket is finite for $y\rightarrow 1$, since $\mathfrak{F}$ stays finite if its argument tends to $\pm\infty$ (compare eqs.\,\eqref{eq:Ffrak limiting behavior +infty} and \eqref{eq:Ffrak limiting behavior -infty}). Hence, analogous to eq.\,\eqref{eq:D0st behavior}
\begin{align}
D_0(t,u,Q^2)=(1-y)^{-1-\eps} A_{tu}(y)\,.
\label{eq:D0tu behavior}
\end{align}
For $D(s,u,Q^2)$ we obtain
\begin{align}
D(s,u,Q^2)&\sim\frac{1}{su}\left|\frac{t \mu^2}{s u}\right|^\eps \left[\mathfrak{F}\!\left(\eps;-\frac{t}{s}\right)+\e^{\iu \pi\eps}\,\mathfrak{F}\!\left(\eps;-\frac{t}{u}\right)-\e^{\iu \pi\eps}\,\mathfrak{F}\!\left(\eps;-\frac{t Q^2}{s u}\right)\right]
\nonumber\\
&\sim (1-y)^{\eps}\left[\mathfrak{F}\!\left(\eps;(1-z)(1-y)\right)+\e^{\iu \pi\eps}\,\mathfrak{F}\!\left(\eps;-\frac{1-y}{y}\right)\right.\nonumber\\
&\hphantom{\sim (1-y)^{-1-\eps}}\,\,\left.-\e^{\iu \pi\eps}\,\mathfrak{F}\!\left(\eps;-\frac{(1-y) z}{y}\right)\right].
\end{align}
In the limit $y\rightarrow 1$ the arguments of all three $\mathfrak{F}$ functions become zero. 
From eq.\,\eqref{eq:Ffrak limiting behavior 0} we infer that for $\tilde{\mathfrak{F}}(\eps;x)=\mathfrak{F}(\eps;x)-|x|^\eps$ the limit $\lim_{x\rightarrow 0}\frac{\tilde{\mathfrak{F}}(\eps;x)}{x}$ is finite order by order in $\eps$.
Therefore, we can write
\begin{align}
&D(s,u,Q^2)\sim (1-z)^{-\eps}+\e^{\iu \pi\eps}y^\eps(1-z^{-\eps})\nonumber\\
&\quad+(1-y)^\eps \left[\tilde{\mathfrak{F}}\!\left(\eps;(1-z)(1-y)\right)+\e^{\iu \pi\eps}\,\tilde{\mathfrak{F}}\!\left(\eps;-\frac{1-y}{y}\right)-\e^{\iu \pi\eps}\,\tilde{\mathfrak{F}}\!\left(\eps;-\frac{(1-y) z}{y}\right)\right].
\end{align}
Note that the term in square brackets stays finite in the limit $y\rightarrow 1$ even when divided by $(1-y)$ and hence
\begin{align}
D(s,u,Q^2)&=A_{su}(y)+(1-y)^{1+\eps} B_{su}(y)\,,
\label{eq:D0su behavior}
\end{align}
with both $A_{su}(y)$ and $B_{su}(y)$ finite in the limit $y\rightarrow 1$ order by order in $\eps$. 

To convert the singular behavior for $y\rightarrow 1$ into poles in $\eps$, which is required to combine real and virtual corrections ensuring cancellation of infrared poles, we use the expansion (see e.g. eq.\,(110) in \citep{Altarelli1979} or eq.\,(69) in \citep{Campbell2017})
\begin{align}
(1-y)^{-1-k\eps}=-\frac{1}{k\eps}\,\delta(1-y)+\sum_{n=0}^\infty\frac{(-k\eps)^n}{n!}\left[\frac{\ln^n(1-y)}{1-y}\right]_+\,,
\label{eq:Plus distribution expansion}
\end{align}
where the plus distribution is defined by
\begin{align}
\int_0^1\dx y\,f(y)\,[g(y)]_+\,\equiv\int_0^1\dx y\,[f(y)-f(1)]\,g(y)\,.
\end{align}
Using eqs.\,\eqref{eq:D0st behavior}, \eqref{eq:D0tu behavior}, and \eqref{eq:D0su behavior}, we find for the four types of contributions
\begin{itemize}
\item[(i)] $(1-y)^{-\eps}D_0(s,t,Q^2)=\left(-\frac{1}{2\eps}\,\delta(1-y)+\sum_{n=0}^\infty\frac{(-2\eps)^n}{n!}\left[\frac{\ln^n(1-y)}{1-y}\right]_+\right)A_{st}(y)\,f_1(y)$\,,
\item[(ii)] $(1-y)^{-\eps}D_0(s,u,Q^2)=\left(\sum_{n=0}^\infty \frac{(-\eps)^n}{n!}\ln^n(1-y)\right)(A_{su}(y)+(1-y)B_{su}(y))\,f_2(y)$\,,
\item[(iii)] $(1-y)^{-1-\eps}D_0(s,t,Q^2)=\left(-\frac{1}{\eps}\,\delta(1-y)+\sum_{n=0}^\infty\frac{(-\eps)^n}{n!}\left[\frac{\ln^n(1-y)}{1-y}\right]_+\right)A_{su}(y)\,f_3(y)$\\\phantom{$(1-y)^{-1-\eps}D_0(s,t,Q^2)=$}\vphantom{$\frac{1}{1}$}$+B_{su}(y)\,f_3(y)$\,,
\item[(iv)]$(1-y)^{-\eps}D_0(t,u,Q^2)=\left(-\frac{1}{2\eps}\,\delta(1-y)+\sum_{n=0}^\infty\frac{(-2\eps)^n}{n!}\left[\frac{\ln^n(1-y)}{1-y}\right]_+\right)A_{tu}(y)\,f_4(y)$\,.
\end{itemize}
We see that the order $\eps^1$ of the box integral contributes to the order $\eps^0$ of the full result, since the $\frac{1}{\eps}$-poles generated by eq.\,\eqref{eq:Plus distribution expansion} multiply the order $\eps^1$ of the functions $A_{st}(y)$, $A_{su}(y)$, and $A_{tu}(y)$. Hence, it is required to perform the $\eps$-expansion of the box integral to one order higher than needed if no regularization would be required.
In the case of multiple regularizations, e.g. for $y,z\rightarrow 1$ as in eq.\,(111) of \citep{Altarelli1979}, multiple poles multiply the box integral, requiring knowledge of the box integral to even higher orders in $\eps$.
Furthermore it was crucial to sum up any terms of the box integral singular in the kinematic limit to all orders, i.e. summing up terms of the form $\eps^n\ln^n(1-y)$ to $(1-y)^{-\eps}$. This changed $\frac{1}{\eps}$ to $\frac{1}{2\eps}$ in (i) and (iv).

%%%%%%%%%%%%%%%%%%%%%%%%%%%%%%%%%%%%%%%%%%%%%%%%%%%%%%%%%%%%%%%%%%%%%%%%%%%%%%%%%%%%%%%
%
% The bibliography begins here
%
%%%%%%%%%%%%%%%%%%%%%%%%%%%%%%%%%%%%%%%%%%%%%%%%%%%%%%%%%%%%%%%%%%%%%%%%%%%%%%%%%%%%%%%
\bibliography{Bibliography_scalar_box_integral}
\bibliographystyle{JHEP}
\end{document}